%% file: main.tex
\documentclass{aa}
\usepackage{graphicx}
\usepackage[mathscr]{euscript}
\usepackage{natbib}
\bibpunct{(}{)}{;}{a}{}{,} 
\usepackage{mathrsfs}
\usepackage{bm}
\usepackage{wasysym}
\usepackage{supertabular}
\usepackage{color}
\usepackage{soul}
\usepackage{txfonts}
\usepackage{hyperref}
 \hypersetup{draft}
\begin{document}

   \title{J-PLUS: Identification of low-metallicity stars with \\ artificial neural networks using SPHINX}
   \titlerunning{J-PLUS: Low-metallicity stars with SPHINX}

   \author{D. D.~Whitten\inst{1}\fnmsep\inst{2}
          \and
          %%%%% MAIN CONTRIBUTORS
   		  V. M. Placco\inst{1}\fnmsep\inst{2}%
          \and
          T. C. Beers\inst{1}\fnmsep\inst{2}%
          \and
          A.~L.~Chies-Santos\inst{3}%
          \and
          C.~Bonatto\inst{3}%
          \and
          J.~Varela\inst{4}%
          \and
          D.~Crist\'obal-Hornillos\inst{4}%      %    
          \and
          A.~Ederoclite\inst{4}%
		  \and
          T. Masseron\inst{5,6}
          \and
          Y.~S.~Lee\inst{7}   %   
          \and
		  %%%%% PEOPLE WHO EDITED
          S.~Akras\inst{8}
          \and
          M. Borges Fernandes\inst{8}
          \and
          J.~A.~Caballero\inst{9} % 
          \and
          A.~J.~Cenarro\inst{4}
          \and
          P.~Coelho\inst{10}%% 
          \and
          M.~V.~Costa-Duarte \inst{10}
          \and
          S.~Daflon\inst{8}
          \and
          R.~A.~Dupke\inst{8,13,14}%%%%%%%%CHECKED
          \and
          R.~Lopes~de~Oliveira\inst{15,16, 10, 8}
          \and
          C.~L\'opez-Sanjuan\inst{4}
          \and
          A.~Mar\' \i n-Franch\inst{4}
          \and
          C.~Mendes~de~Oliveira\inst{10}
          \and
          M.~Moles\inst{12}
          \and
          A.~A.~Orsi\inst{12}
          \and
          S.~Rossi\inst{10}
          %%%%% BUILDERS
          \and
          L.~Sodr\'{e}\inst{10}
          \and
		  H.~V\'azquez Rami\'{o}\inst{12}%
          }

	\authorrunning{Devin D. Whitten}
    
   \institute{Department of Physics, University of Notre Dame, Notre Dame, IN. 46556, USA\\
              \email{dwhitten@nd.edu}
         \and
             JINA Center for the Evolution of the Elements (JINA-CEE), USA
         \and
         	Departamento de Astronomia, Instituto de F\'isica, Universidade Federal do Rio Grande do Sul, Porto Alegre, RS, Brazil 
         \and
         	Centro de Estudios de F\'isica del Cosmos de Arag\'on (CEFCA) - Unidad Asociada al CSIC, Plaza San Juan, Planta 2, E-44001, Teruel, Spain
         \and
         Instituto de Astrof\'isica de Canarias, E-38205 La Laguna, Tenerife,Spain
         \and
         Departamento de Astrof\'isica, Universidad de La Laguna, E-38206 La Laguna, Tenerife, Spain
         \and
         Department of Astronomy and Space Science, Chungnam National University, Daejon 34134, Korea
         \and
         Observat\'orio Nacional - MCTIC (ON), Rua Gal. Jos\'e Cristino 77, São Crist\'ov\~ao, 20921-400, Rio de Janeiro, Brazil
         \and
         Centro de Astrobiolog\'ia (CSIC-INTA), ESAC, Camino Bajo del Castillo S/N, E-28692 Villanueva de la Ca\~nada Madrid, Spain
         \and
         Instituto de Astronomia, Geof\'{i}sica e Ci\^{e}ncias Atmosf\'{e}ricas (IAG), Universidade de S\~{a}o Paulo (USP), Rua do Mat\~{a}o 1226, C. Universit\'{a}ria, S\~{a}o Paulo, 05508-090, Brazil
         \and
         Departamento de F\'{i}sica, Universidade Federal de Sergipe (UFS), Av. Marechal Rondon, S/N, 49000-000 S\~{a}o Crist\'{o}v\~{a}o, SE, Brazil
         \and
         Centro de Estudios de F\'{i}sica del Cosmos de Arag\'{o}n (CEFCA), Plaza San Juan, 1, E-44001, Teruel, Spain
         %Observat\'orio Nacional, Rua Gal. Jos\'{e}, Cristino 77, S\~{a}o Crist\'{o}v\~{a}o, Rio de Janeiro, RJ, Brazil
         %\and
         %Department of Astronomy and Space Science, Chungnam National University Daejeon 34134, Korea
         \and
         University of Michigan, Dept. Astronomy, 1085 S. University Ann Arbor, MI 48109, USA
         \and
         University of Alabama, Dept. of Phys. \& Astronomy, Gallalee Hall, Tuscaloosa, AL 35401, USA
         \and
         X-ray Astrophysics Laboratory, NASA Goddard Space Flight Center, Greenbelt, MD 20771, USA
         \and
         Department of Physics, University of Maryland, Baltimore County, 1000 Hilltop Circle, Baltimore, MD 21250, USA  
        }
   \date{Received 15 May 2018}

% \abstract{}{}{}{}{} 

  \abstract
{We present a new methodology for the estimation of stellar atmospheric parameters from narrow- and intermediate-band photometry of the Javalambre Photometric Local Universe Survey (J-PLUS), and propose a method for target pre-selection of low-metallicity stars for follow-up spectroscopic studies. Photometric metallicity estimates for stars in the globular cluster M15 are determined using this method.}
{By development of a neural-network-based photometry pipeline, we aim to produce estimates of effective temperature, $T_{\rm eff}$, and metallicity, [Fe/H], for a large subset of stars in the J-PLUS footprint.}
{The Stellar Photometric Index Network Explorer, SPHINX, is developed to produce estimates of $T_{\rm eff}$ and [Fe/H], after training on a combination of J-PLUS photometric inputs and synthetic magnitudes computed for medium-resolution ($R \sim$2000) spectra of the Sloan Digital Sky Survey. This methodology is applied to J-PLUS photometry of the globular cluster M15.}
{Effective temperature estimates made with J-PLUS Early Data Release photometry exhibit low scatter, $\sigma(T_{\rm eff})$ = 91\,K, over the temperature range 4500 $< T_{\rm eff}$ (K) $<$ 8500.
For stars from the J-PLUS First Data Release with 4500 $< T_{\rm eff}$ (K) $<$ 6200, $85 \pm 3$\,\% of stars known to have [Fe/H] $<-2.0$ are recovered by SPHINX. A mean metallicity of [Fe/H]=$-2.32\pm 0.01$, with a residual spread of 0.3\,dex, is determined for M15 using J-PLUS photometry of 664 likely cluster members.}
{We confirm the performance of SPHINX within the ranges specified, and verify its utility as a stand-alone tool for photometric estimation of effective temperature and metallicity, and for pre-selection of metal-poor spectroscopic targets.}


   \maketitle
%
%_________________________________________________________________________________

\section{Introduction}

\par The chemical properties of individual stars in the Milky Way are crucial in order to develop an understanding of our Galaxy's chemical evolution and assembly history. In particular, the metallicity distribution function of Galactic halo stars is among the most important observational constraints for cosmological models \citep{Beers:2005, Salvadori:2010}. The comparatively rare stars with metallicity below 1\% of the Solar value -- described in terms of their metal abundance, very metal-poor (VMP; [Fe/H]\footnote{[A/B]$\equiv \log_{10}(N_{\rm A}/N_{\rm B})_* - \log_{10}(N_{\rm A}/N_{\rm B})_{\odot}$} $< -2.0$), extremely metal-poor (EMP; [Fe/H] $< -3.0$), and ultra metal-poor (UMP; [Fe/H] $< -4.0$) -- are expected to include the earliest generations of stars to have formed since the Big Bang. With the exception of mass-transfer binaries and highly evolved late-type giants, these ancient stars retain the chemical signature of their natal environments. Measurement of the chemical abundances of the earliest stars thereby provides a means to study the nucleosynthetic pathways and astrophysical mechanisms that were in operation during the first generations of stars born in the early Universe. One example is the most iron-poor star presently known, 
SMSS~J031300.36-670839.3 ([Fe/H] $\le -7.1$), from which the carbon ([C/Fe]) and [Mg/Ca] abundance ratios are believed to have originated from a single explosion of a metal-free $\sim 60$\,M$_{\astrosun}$-mass star (\citealt{Keller:2014}).

\begin{figure*}
	\centering
	\includegraphics[trim={2.0cm 0.0cm 2.0cm 0.0cm}, width=\textwidth]{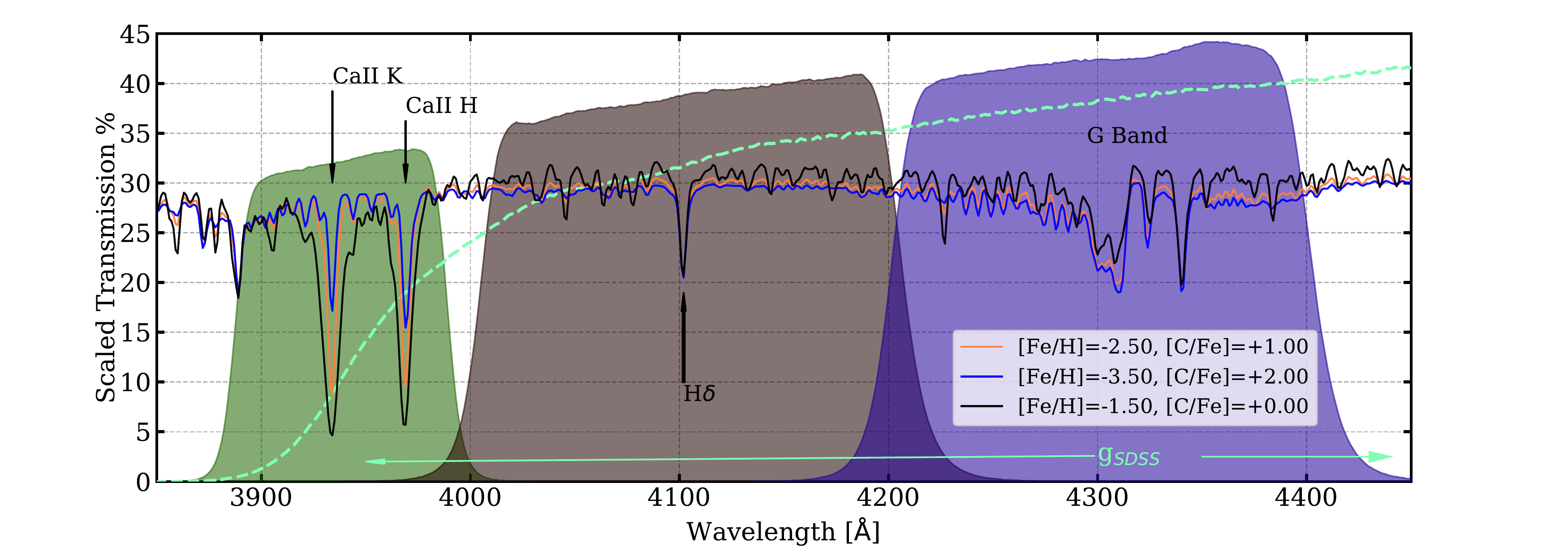}
	\caption{Synthetic spectra of varying abundances are shown plotted over the J-PLUS filters $J0395$, $J0410$, and $J0430$. The $J0395$ filter is optimally centered on the \ion{Ca}{II} H \& K lines, which can be used as indicators of the overall stellar metallicity. The temperature-sensitive H$\delta$ line is apparent in the F410 filter. The $J0430$ filter captures the CH $G$-band, which serves as a carbon indicator.}
	\label{fig:zoom}
\end{figure*}

\par A substantial fraction of VMP stars exhibit large enhancements of their [C/Fe] ratios, increasing rapidly with declining [Fe/H] (see Yoon et al. 2018 and references therein); these are known collectively as carbon-enhanced metal-poor (CEMP) stars \citep{Beers:2005}. Guided by the work of \citet{Spite:2013} and \citet{Bonifacio:2015}, \citet{Yoon:2016} explored the complex morphology of CEMP stars in the $A$(C)-[Fe/H] space\footnote{$A$(X) = log $\epsilon$(X) = log(N$_{\rm X}$/N$_{\rm H}$) + 12, where $\rm X$ represents a given element.}. The so-called Yoon-Beers diagram (Fig. 1 of Yoon et al.) provides evidence for multiple progenitors and/or environments in which different sub-classes of CEMP stars are found. 
The identification and study of significantly larger samples of CEMP stars is crucial for future studies, as only $300+$ of these stars with available high-resolution spectroscopy are currently known.

Obtaining chemical abundances and overall estimates of metallicity is a costly endeavor, however, requiring pre-selection and follow-up spectroscopic analysis for confirmation. While there are now tens of thousands of VMP stars with well-measured (medium-resolution) spectroscopic metallicities, the numbers of known EMP and UMP stars are considerably smaller; in particular, $\sim$30 UMP stars have been discovered to date (the compilation as given in \citealt{Placco:2015a, Placco:2016b}, and \citealt{Abohalima:2017, Starkenburg:2018, Frebel:2018, Aguado:2018}).

\subsection{Photometric methods}

Wide-field photometry offers an alternative means to probe the chemical characteristics of stars, and presents a method for pre-selection of targets based on their colors. Using broad-band photometry from the Sloan Digital Sky Survey (SDSS; \citealt{York:2000}), \citet{Ivezic:2008a} developed a methodology for estimating metallicity for F- and G-type main-sequence stars ($5000 < T_{\rm eff}$\,(K) $<7000$, $3 < \log{g} < 5$) using polynomial regressions based on de-reddened $u-g$ and $g-r$ colors. This approach was found to be effective down to [Fe/H] $\sim -2.0$.

Later, \citet{An:2013} and \citet{An:2015} used fiducial isochrone fits to SDSS \textit{ugriz} photometry to extend metallicity determinations for main-sequence stars with broad-band photometry down to at least [Fe/H] $\sim$ $-2.5$. This new threshold is below the peak in the metallicity distribution function of outer-halo population stars ([Fe/H]$ =-2.2$; \citealt{Carollo:2007,Carollo:2010}), leading them to conclude that $\sim$35–-55\% of local halo stars belong to this population. Other notable moderate- and narrow-band photometric metallicity estimation techniques have existed for many years, for instance, the Washington \citep{Canterna:1976} and Str\"omgren \citep{Stromgren:1963, Stromgren:1964} systems. However, no wide-field large sky-coverage surveys in these systems have been carried out to date. The Canada-France Imaging Survey, which will cover 10,000\,deg$^2$ of the northern sky with a $u$-band, demonstrated metallicity sensitivity of $\sigma$[Fe/H] = 0.2\,dex down to [Fe/H] $\sim-2.5$ for dwarf stars (3 < log$g$ < 5) when combined with SDSS and PS1 photometry \citep{Ibata:2017a, Ibata:2017b}.

Pre-selection of targets for spectroscopic follow-up can dramatically increase the success rate for the identification of large numbers of metal-poor stars. For example, using all-sky APASS optical, 2MASS near-infrared, and {\em WISE} mid-infrared photometry, \citet{Schlaufman:2014} developed an efficient method for selecting bright ($V <$ 14) metal-poor candidates based on their lack of molecular absorption near 4.6\,$\mu$m. This effect was demonstrated to be present in atmospheres of all surface gravities in the effective temperature range $4500 \lesssim T_{\rm eff}$ (K) $ \lesssim 5500$. Of these targets, 32.5\% were found to have $-3.0 < $ [Fe/H] $< -2.0$. The identification of bright metal-poor candidates is of great benefit to high-resolution follow-up observations, for which the acquisition of high-resolution spectra with high signal-to-noise can become prohibitively expensive in terms of telescope time.

New multi-band photometric surveys build upon previous broad-band photometric metallicity determinations by implementation of one or more narrow-band filters across the optical and near-infrared spectrum to target key stellar absorption features. The SkyMapper Southern Survey \citep{Keller:2007} makes use of a 310\,{\AA} $v$-band filter, which covers (but is not centered on) the \ion{Ca}{II} K line, along with SDSS-like $ugriz$ filters, to survey much of the entire Southern Hemisphere sky. Other recent ongoing surveys, such as Pristine \citep{Starkenburg:2017}, are employing a narrow-band ($\sim$100\,\AA) filter centered on the \ion{Ca}{II} H \& K lines, a technique pioneered by \citet{Anthony-Twarog:1991}. When used in conjunction with pre-existing SDSS or Pan-STARRs \citep{Tonry:2012} broad-band filter photometry, they obtain improved photometric metallicity estimates for numerous stars over large swaths of sky. As described by \citet{Starkenburg:2017}, Pristine's resulting success rate for recovering EMP stars is 26\%, with 80\% of the remaining candidates being VMP stars. When combined with SDSS photometry, Pristine has demonstrated an accuracy of $\sim$0.2\,dex down to [Fe/H] $< -$3.0. The use of \ion{Ca}{II} H \& K photometry from Pristine has already led to the discovery of a [Fe/H]$<-4.66$ star, Pristine\_221.8781+9.7844 \citep{Starkenburg:2018}.

%_________________________________________________________________
% Javalambre Photometric Local Universe Survey
%______________________________________________________________
\subsection{The Javalambre Photometric Local Universe Survey}

Located at the Observatorio Astrof\'{i}sico de Javalambre (OAJ, Teruel, Spain), the Javalambre Auxiliary Survey Telescope (JAST/T80) is a 83\,cm telescope that is currently carrying out the Javalambre Photometric Local Universe Survey (J-PLUS; \citealt{Cenarro:2018}). A twin of JAST/T80, based at the Cerro Tololo InterAmerican Observatory (CTIO, Chile), is executing the Southern Photometric Local Universe Survey (S-PLUS; Mendes de Oliveira et al. 2018), a Southern Hemisphere counterpart of J-PLUS. 

J-PLUS is the first large-sky survey conducted at the OAJ, and was initially conceived to aid with photometric calibration of the upcoming Javalambre Physics of the Accelerating Universe Survey (J-PAS; \citealt{Benitez:2014}), to be executed as well at the OAJ. While J-PAS is motivated by cosmological goals and, hence, has formally stronger requirements than J-PLUS in terms of photometric depth and number of narrow-band filters (see \citet{Benitez:2014} for details), J-PLUS has scientific goals that are largely (but not exclusively) related to the science of the Milky Way Halo and local Universe studies.
The seven narrow-band ($\sim$100\,\AA) filters of J-PLUS ($J0378$, $J0395$, $J0410$, $J0430$, $J0450$, $J0515$, $J0660$, $J0861$) are specifically designed to detect a variety of absorption features across the optical spectrum (\citet{Cenarro:2018} for details).
The filter system also hosts traditional $ugriz$ bands, similar to those of SDSS, but with a more stable $u$-band response (Cenarro et al. 2018). Fig. \ref{fig:zoom} depicts the location of some filters of special interest -- $J0395$, $J0410$, and $J0430$, from left to right -- along with the corresponding spectral features, illustrated with synthetic spectra.

These filters provide the sensitivity necessary to detect absorption features such as the \ion{Ca}{II} H \& K lines, and the CH $G$-band, but mapping their behavior to estimate metallicity, [Fe/H], and the carbon-to-iron abundance ratio, [C/Fe], presents a highly degenerate problem. At temperatures above $\sim5750$\,K, the CH molecule begins to dissociate, while the H$\gamma$ line broadens to the extent of disrupting the feature. The \ion{Ca}{II} H \& K lines also exhibit a strong temperature dependence. In addition, many lines respond to varying surface gravity due to Stark pressure broadening. A robust technique capable of unraveling the complex behavior of the parameter space explored by these filters for estimates of stellar parameters is thus required.

\subsection{Artificial Neural Networks}

Machine learning tools -- and in particular, Artificial Neural Networks (ANN) -- have a long history \citep{Rosenblatt:1958}, but have found success in data-driven astronomical applications. Once adequately trained, these networks are powerful statistical pattern recognition tools capable of modeling highly complex behavior, through implementation of non-linear activation functions according to a user-specified architecture. \citet{Fabbro:2018} applied their deep convolutional neural network, \texttt{StarNet}, to SDSS-III APOGEE \citep{Eisenstein:2011} spectra for determination of effective temperature, surface gravity, and metallicity, with precision and accuracy similar to that of the APOGEE pipeline. Using APOGEE DR14 for validation, \citet{Ting:2018} applied a neural network function in their method, \textit{The Payne}, to interpolate training sets of physical \textit{ab initio} spectral models, and produced estimates of stellar parameters and 15 element abundances.

Deep convolutional networks have also been used to determine the evolutionary states of red giants from astroseismology \citep{Hon:2017}. Using frequency power spectra from the {\em Kepler} mission \citep{Borucki:2010}, \citet{Hon:2018} classified 426 red giants as red giant branch or helium-core burning stars.
Neural networks have been employed for classification-based problems in astronomy as well, such as star-galaxy classification \citep{Kim:2017} and spectral classification \citep{Kheirdastan:2016}.

In this paper, the utility of ANNs for estimates of effective temperature and metallicity is explored. We present the Stellar Photometric Index Network Explorer (SPHINX\footnote{\texttt{https://github.com/DevinWhitten/SPHINX}}), a software package based on ANN estimation of stellar parameters ($T_{\rm eff}$ and [Fe/H]) from the J-PLUS mixed-bandwidth photometry. All training catalogs and parameter determination routines used in the text are provided.
The databases and ANN used in the development and operation of SPHINX are described in Sections~\ref{databases} and \ref{network}. The basic structure is outlined in Section~\ref{SPHINX}. In Section~\ref{teff_section}, the training process and results of effective temperature determinations are described, followed by the results of the metallicity determinations in Section~\ref{feh_section}.
An application of SPHINX to a case study of the metal-poor globular cluster M15 is conducted in Section~\ref{M15_section}. A preliminary investigation of carbon sensitivity within the J-PLUS filter system is described in Section~\ref{carbon_section}, and a concluding discussion of results and future applications is given in Section~\ref{conclusion}. 
%
%______________________________________________________________

\section{Databases}\label{databases}
\begin{figure*}
	\centering
	\includegraphics[trim={1.0cm 0.00cm 0.0cm 0.00cm},clip,width=\textwidth]{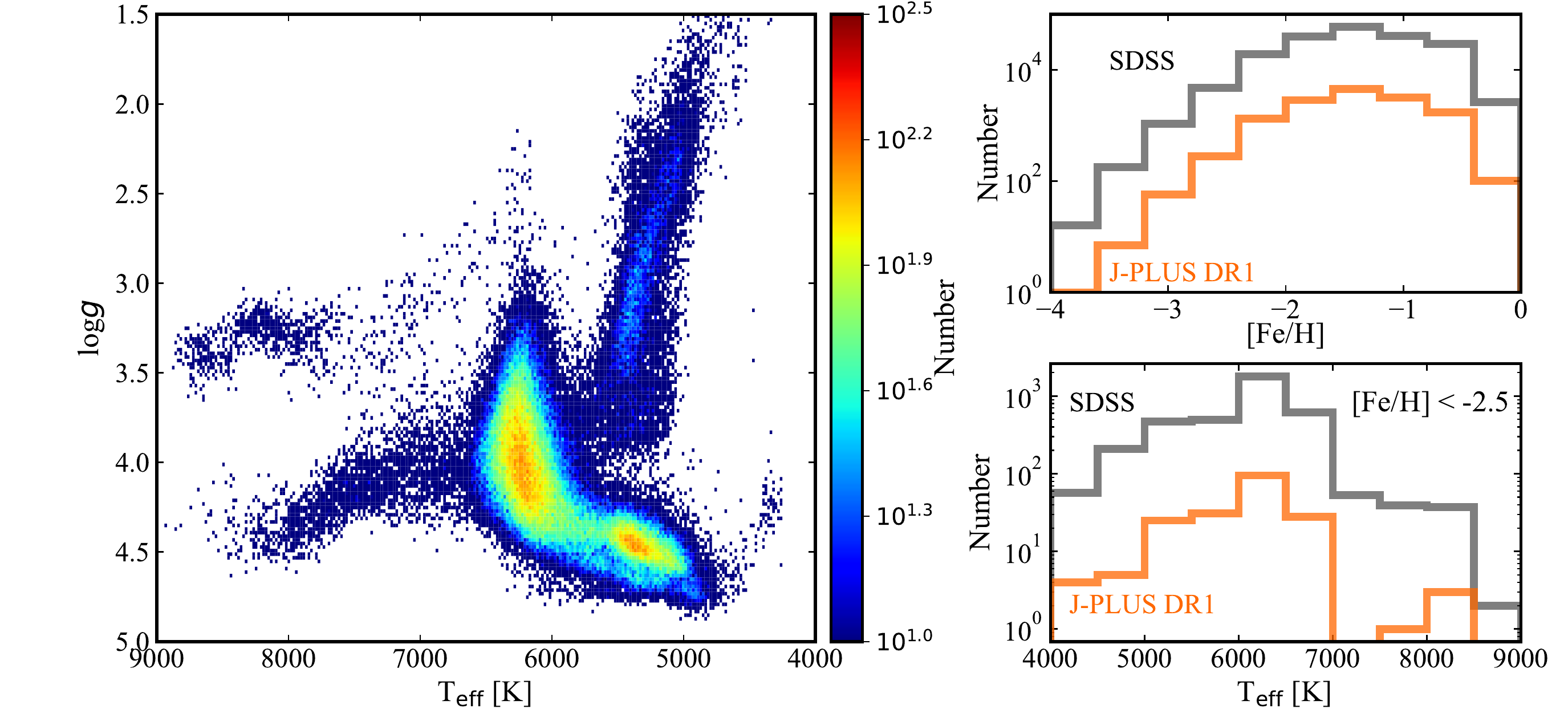}
	\caption{\textit{Left:} Hess diagram of the SDSS Reservoir available for training of the photometric ANN pipeline. \textit{Top Right:} Metallicity distribution of the SDSS Reservoir and the subset with J-PLUS DR1 photometry. \textit{Bottom Right:} Distribution of effective temperatures for the subset of the SDSS Reservoir with [Fe/H] < -2.5.}
	\label{fig:SEGUE}
\end{figure*}
\subsection{The J-PLUS Early and First Data Releases}

Calibration and testing of the network's inputs are performed in part with the J-PLUS data set. The J-PLUS First Data Release\footnote{\url{j-plus.es/datareleases/data\_release\_dr1}} (hereafter DR1) consists of 511 pointings collected from November 2015 to January 2018 with JAST/T80 in all twelve optical bands described above, and covers approximately 1022\,deg$^2$. The reduction and photometric calibration of the J-PLUS DR1, as well as the limiting magnitudes in the twelve bands, are presented in \citet{Cenarro:2018}. A representative subset of the DR1, the J-PLUS Early Data Release (EDR), comprises 18 pointings (36\,deg$^2$), and is publicly available\footnote{\url{j-plus.es/datareleases/early_data_release}}. In addition to the present paper, the J-PLUS EDR and science verification data have been used to refine the membership in the nearby galaxy clusters Abell 2589 and Abell 2593 \citep{Molino:2018}, analyze the globular cluster M15 \citep{Bonatto:2018}, study the H$\alpha$ emission \citep{Logronho:2018} and the stellar populations \citep{Sanroman:2018} of several local galaxies, and compute the stellar and galaxy number counts up to $r = 21$ \citep{Lopez:2018}.

Selections of the 6\,arcsec aperture photometry are made from the \texttt{MagABDualObj} catalogs for each release. To limit contamination from non-stellar type objects, we remove extended sources using the stellarity index given by \texttt{SExtractor}, \texttt{CLASS\_STAR} $\geq$ 0.92 \citep{Bertin:1996, Lopez:2018}. This morphological classifier is itself based on a neural network, taking into account the pixel scale of the image and full width at half maximum.

From these selections, we obtained 3,132,543 and 78,329 sources for DR1 and EDR, respectively. Minor variations in the 6\,arcsec aperture photometry were present, tile by tile, in EDR. Corrections were made, based on a stellar-locus regression technique provided by Crist\'obal-Hornillos et al. (priv. comm.). For training and validation of SPHINX, these catalogs were 
cross-matched, using a 1.0\,arcsec search radius, with stars from the SDSS Reservoir (described in Section \ref{SDSS_reservoir}) with available stellar atmospheric-parameter estimates. This search radius was varied to 3.0\,arcsec, and no new stars were obtained.

Interstellar extinction is expected to be small in EDR and DR1. However, application of extinction corrections is still expected to improve the performance of the photometric ANNs. One method to determine reddening coefficients for J-PLUS photometry is to convolve the filter-response functions corresponding to each of the J-PLUS filters with the reddening law from \citet{Fitzpatrick:1999}. The result is a transmission-weighted sum of the wavelength-dependent extinction contribution within the filter bandpass. These values are provided in Table \ref{extinction_table}. The effective wavelength, $\lambda_{\rm eff}$, is determined from a transmission-weighted average of the wavelength for each filter.

\begin{table}      % is used to refer this table in the text
% title of Table

    \caption{Reddening coefficients for the J-PLUS photometric system.\label{extinction_table}}
	\centering                          % used for centering table
	\begin{tabular}{c c c}        % centered columns (4 columns)
		\hline\hline                 % inserts double horizontal lines
		Filter & $\lambda_{\rm eff}$ & A($\lambda$)/$E(B-V)$ \\    % table heading 
		(name) & (\AA) &  \\
		\hline                        % inserts single horizontal line
        \textit{u} & 3536 & 4.295 \\
		\textit{J}0378 & 3782 & 4.013 \\      % inserting body of the table
		\textit{J}0395 & 3939 & 3.849  \\
		\textit{J}0410 & 4108 & 3.685 \\
		\textit{J}0430 & 4303 & 3.506 \\
        \textit{g} & 4810 & 3.120 \\
		\textit{J}0515 & 5141 & 2.864 \\ 
        \textit{r} & 6271 & 2.235 \\
		\textit{J}0660 & 6604 & 2.070 \\
        \textit{i} & 7669 & 1.668 \\
		\textit{J}0861 & 8611 & 1.377 \\
		\textit{z} & 8979 & 1.290 \\
		
		\hline         \\                          %inserts single line
	
    \end{tabular}

\end{table}

To estimate the line-of-sight reddening, $E(B-V)$, for our sources, we use \citet{Schlafly:2011}. Corrections for reddening estimates for the J-PLUS photometry indeed improve the synthetic magnitude calibrations (Section \ref{synth_section}). However, these corrections are subject to some limitations, particularly in that this extinction map is only two-dimensional. In future work, we will utilize three-dimensional estimates of dust extinction \citep[e.g.,][]{Green:2018}, based on parallaxes from {\em Gaia} Data Release 2 \citep{Gaia:2016}. In this work, we implement a hard extinction selection, corresponding to $E(B-V)$ < 0.05\,dex, so that more reddened sources are not admitted to our samples.

\subsection{The SDSS Reservoir}\label{SDSS_reservoir}

Three separate spectroscopic campaigns comprise the training databases for SPHINX, hereafter called the SDSS Reservoir. The first, SDSS's Baryon Oscillation Spectroscopic Survey (BOSS; \citealt{BOSS}), obtained over 500,000 stellar spectra (including repeated stars) across a wavelength of 3,600 - 10,000 \AA \hspace{0.1cm} (R $\approx$ 2,000) over approximately 10,000\,deg$^2$ of the sky. Our database includes 80,221 stars from this survey, selected for high signal-to-noise and low metallicity ([Fe/H] $< -0.5$). The Sloan Extension for Galactic Understanding and Exploration (SEGUE; \citealt{Yanni:2009}) took place in two phases, SEGUE-1 and SEGUE-2, and covered a combined area of 2,755\,deg$^2$. We make use of 147,811 of the $\sim$350,000 spectra from this sample. Finally, we include 74,572 spectra from the component of SDSS-I known as the Legacy Survey, provided by SDSS DR7 \citep{Abazajian:2009}. This survey covered primarily $\sim$7,500\,deg$^2$ of the North Galactic Cap, with additional stripes in the South Galactic Cap amounting to $\sim$740\,deg$^2$.

Stellar parameters for these databases were derived using the SEGUE Stellar Parameter Pipeline (SSPP; \citealt{Lee:2008a,Lee:2008b}). This pipeline made robust determinations of effective temperature ($T_{\rm eff}$), surface gravity ($\log{g}$), metallicity ([Fe/H]), and carbonicity ([C/Fe]) over a temperature range $4000 < T_{\rm eff}$ (K) $ < 8000$. Typical random uncertainties in the effective temperature and metallicity estimates for F- and G-type stars were determined empirically, $\sigma(T_{\rm eff}) \sim$ 130\,K, and $\sigma$([Fe/H])$ \sim$ 0.21\,dex \citep{Prieto:2007}. 
The color-magnitude diagram of stars available for training from the SDSS Reservoir is shown in Fig.~\ref{fig:SEGUE}, along with the effective temperature and metallicity distributions of source matches with J-PLUS DR1. In all three plots, stars were limited to those with $\sigma(T_{\rm eff})<250$\,K, $\sigma$[Fe/H] $<0.3$\,dex, and $S/N>25$, as dictated by the SSPP (where $S/N$ is the signal-to-noise ratio).

\begin{figure*}
	
	\centering
	\includegraphics[trim={0.0cm 1.0cm 0 1.00cm},clip, scale=0.85]{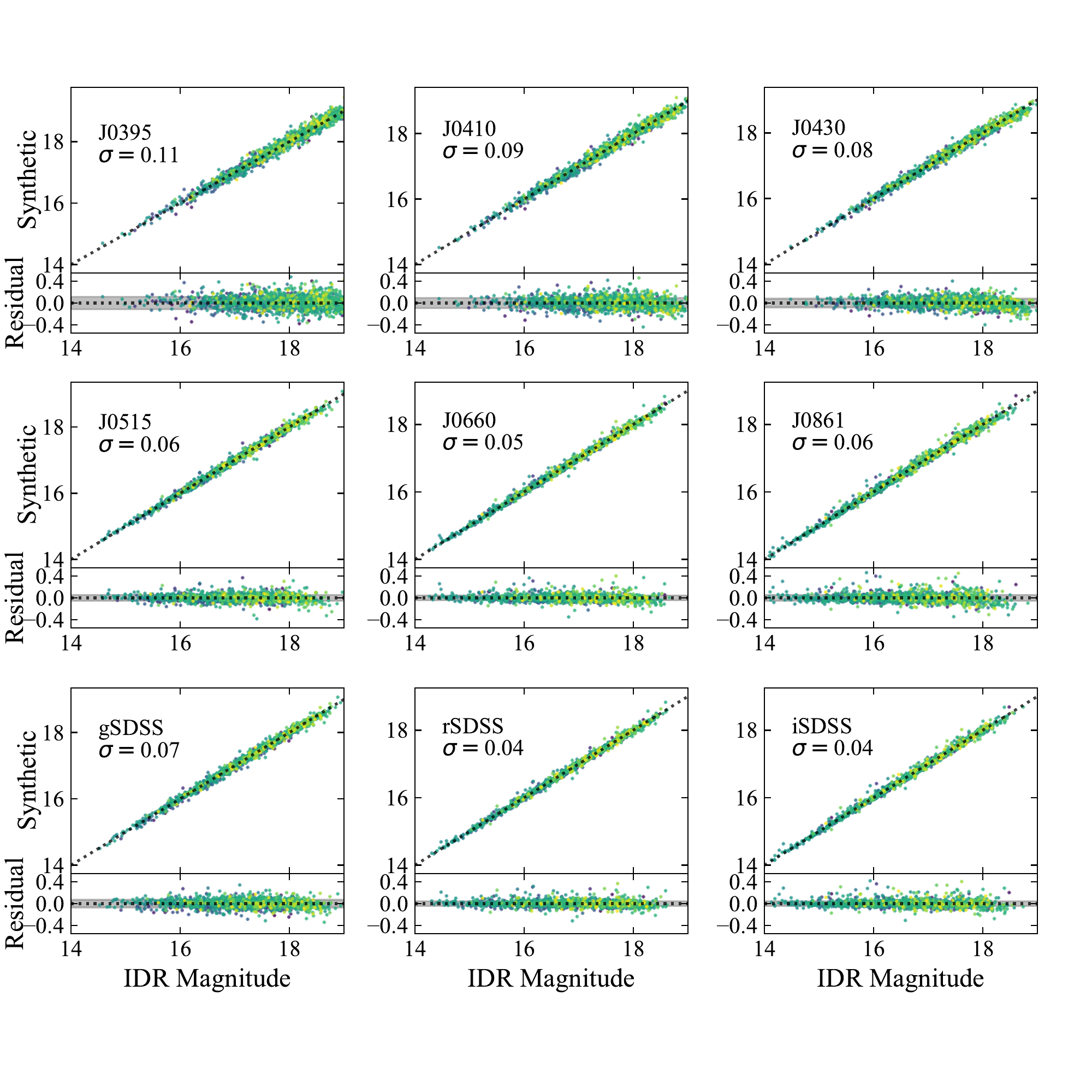}
	\caption{Synthetic magnitudes determined with SDSS Reservoir and J-PLUS filter-response functions are compared with photometry from the J-PLUS First Data Release. Scatter in the calibrations increases for bluer filters, as the uncertainty in the interstellar extinction correction is larger. The colors are indicative of the photometric tile from which the data were obtained, for which no biases are seen.}
	\label{fig:DR1_SEGUE_magnitude}
\end{figure*}

\subsection{Synthetic magnitudes}\label{synth_section}

\begin{figure*}
	\centering
	\includegraphics[trim={0.90cm 1.0cm 0.0cm 0.00cm},clip, width=0.85\linewidth]{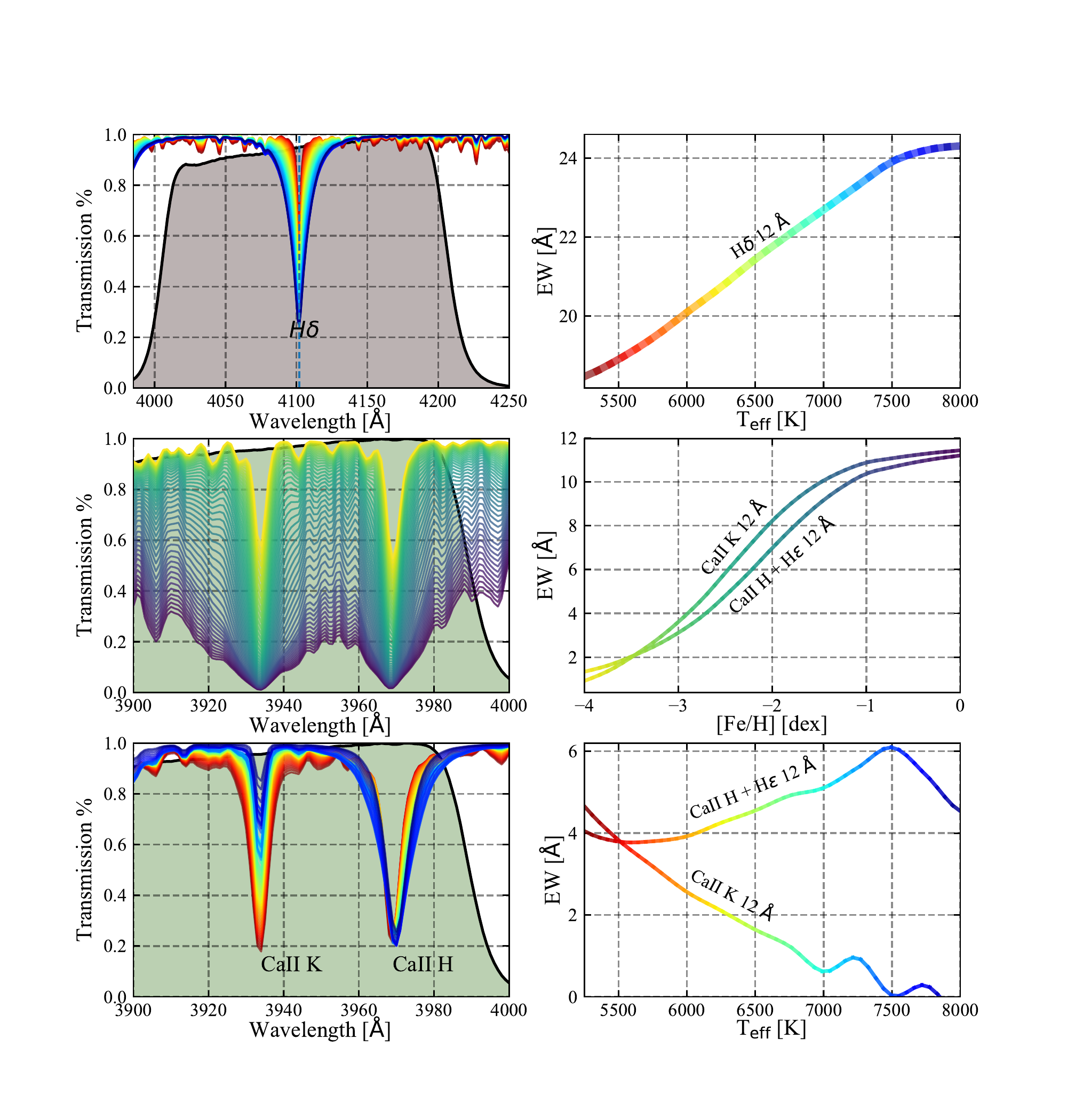}
	\caption{\textit{Top panel}: Temperature sensitivity of the H$\delta$ feature. Over an effective temperature range of 5250 $< T_{\rm eff}$ (K) $< 8000$, we expect an increase in absorption of 31\% for a star of $\log{g}=2.50$, [Fe/H] $=-2.50$, and [C/Fe] $=0.0$. \textit{Middle panel}: The variation in \ion{Ca}{II} H \& K line strengths with increasing metallicity for a star of $T_{\rm eff}=5000$\,K, $\log{g}=1.0$, and [C/Fe] $=0.0$. \textit{Bottom panel}: The \ion{Ca}{II} H \& K line response to increasing effective temperature. The \ion{Ca}{II} H line is nearly superposed with H$\epsilon$.}
	\label{fig:temp_HD}
\end{figure*}

Synthetic photometry was computed for all spectra in the databases described above. This was done initially to provide the photometric inputs required for network training before J-PLUS had accumulated an adequate number of source photometry with EDR and DR1. 
First, filter intensities were computed by taking the inner product of the flux-calibrated spectrum and the J-PLUS filter-response functions. These filter intensities were converted to synthetic magnitudes using the following form of the asinh magnitude, or luptitude \citep{Lupton:1999}:

\begin{equation}
m = (-2.5/\ln{10})\cdot[\sinh^{-1}{\frac{f}{2bf_0}} + \ln{b}].
\end{equation}

\noindent Here, $f$ is the integrated flux through the filter. Appropriate values for the classical zero-point, $f_0$, and the softening parameter, $b$, were determined by calibrating the synthetic intensities to known J-PLUS magnitudes from a cross-match of EDR and DR1 with synthetic magnitudes derived from the flux-calibrated spectra for stars in common with the SDSS Reservoir. This was done using the orthogonal distance regression technique \citep{Boggs:1989}, taking into account error estimates in the native photometry as well as the variances in the flux reported in the medium-resolution SDSS spectra. 

\par The resulting synthetic magnitude calibrations for stars in the SDSS Reservoir are shown in Fig. \ref{fig:DR1_SEGUE_magnitude}. In addition, minor offsets ($\pm0.2$\,mag) were found and removed in a few J-PLUS tiles during the calibration. Residuals in the synthetic magnitude calibrations are moderately dependent on the $S/N$ of the underlying SDSS spectra, particularly for the $J0395$ and $J0410$ filters. Sources with $S/N < 25$ and magnitude uncertainties larger than 0.1\,mag were thus excluded for each calibration. This $S/N$ cut has the effect of preferentially removing calibration stars at fainter magnitudes, where we find that the scatter for each calibration tends to increase. Considering again a critical limit in the scatter of 0.1\,mag, we find limiting magnitude values for each filter, provided in Table~\ref{magnitude_table} along with the standard deviation, $\sigma_m$, of the corresponding calibration. Use of synthetic magnitudes beyond these limits is not recommended, as the calibrations are less reliable. Both uncertainties in the underlying SDSS spectra and the native J-PLUS photometry contribute to the scatter seen in the Fig. \ref{fig:DR1_SEGUE_magnitude}, but if synthetic magnitudes are limited to those with $S/N > 25$, the critical residual value of 0.1\,mag is reached at characteristically brighter values in the native photometric system.

\begin{table}      % is used to refer this table in the text
% title of Table

    \caption{Limiting magnitudes for synthetic calibrations.\label{magnitude_table}}
	\centering                          % used for centering table
	\begin{tabular}{c c c}        % centered columns (4 columns)
		\hline\hline                 % inserts double horizontal lines
		Filter & m$_{lim}$ & $\sigma_m$ \\    % table heading 
		(name) & (mag) &  (mag)\\
		\hline                        % inserts single horizontal line
		\textit{J}0395 & 18.46 & 0.11  \\
		\textit{J}0410 & 18.52 & 0.09 \\
		\textit{J}0430 & 18.69 & 0.08 \\
        \textit{g}     & 18.68 & 0.07 \\
		\textit{J}0515 & 18.79 & 0.05 \\ 
        \textit{r}     & 18.90 & 0.04 \\
		\textit{J}0660 & 18.80 & 0.04 \\
        \textit{i}     & 18.75 & 0.04 \\
		\textit{J}0861 & 18.48 & 0.05 \\
		
		\hline         \\                          %inserts single line
	
    \end{tabular}

\end{table}

As expected, the scatter in the photometric calibration was seen to increase for bluer filters, where CCD response and atmospheric absorption become more problematic. This calibration ensured that synthetic magnitudes were on the same scale as the J-PLUS photometric system, thus both synthetic and native photometry could be used interchangeably within the magnitude and S/N limits specified.

The blue cutoff of the SEGUE spectra ($\sim$ 3700\,\AA) preclude determination of synthetic magnitudes for the $J0378$ and $u$-band filters. The same is true for the $z$-band filter, for which the response curve extends beyond the range of the SEGUE spectra. Consequently, these filters are excluded from the methodology. Once photometry is available for a sufficient number of these stars, construction of synthetic magnitudes for training of the ANNs will not be necessary, and these filters should certainly be included, as they each capture important information about the stellar atmospheres.

\subsection{The synthetic library}

For exploratory and training purposes, we make use of a library of synthetic spectra. To build the synthetic spectra library, we proceed similarly to the approach described by \citet{Lee:2013}. In short, we use a specific grid of model atmospheres computed with the MARCS code \citep{Gustafsson:2008}, which takes into account the impact of the carbon enhancement in the atmosphere. From those model atmospheres we generate the synthetic spectra using the Turbospectrum routine \citep{turbospectrum, Plez:2012}. Compared to  \citet{Lee:2013}, we improve the procedure in several aspects. First, we extend the wavelength range, now covering the entire optical region from 3000\,{\AA} to 10,000\,{\AA}. In addition, we update the linelists, which now include CN, CH, NH, C$_2$, MgH, TiO, ZrO, CaH, VO, and SiH for the molecules, and the ultimate version of VALD3 \citep{VALD3} for the atoms. Finally, we extend the grid to a much larger space of parameters, $3500\le T_{\rm eff}$ (K) $\le 8000$ in increments of 250\,K, $0.0 \le$ $\log{g}$ $\le 5.0$ in increments of 0.5\,dex, $ -4.5 \le$ [Fe/H] $\le +0.5$ in increments of 0.25\,dex, and $-1.5 \le $ [C/Fe] $\le +4.5$ in increments of 0.25\,dex.

Due to the finite resolution of this library, we interpolate when necessary using a 4D cubic spline interpolation routine (Lee, Y.S., priv. comm.).
Training sets can then be generated with stellar parameters distributed anywhere within the parameter regions of interest. A crude flux-calibration is performed on these spectra by convolving each normalized spectrum with a blackbody function corresponding to its effective temperature. This library is primarily used for investigation of the photometric carbon sensitivity, discussed in Section \ref{carbon_section}.

%%%%%%%%%%%%%%%%%%%%%%%%%%%%%%%%%%%%%%%%%%%%%%%%%%%%%%

\subsection{Narrow-band parameter response}

In addition to sampling the overall blackbody function of the spectral-energy distribution (SED), we expect response to a number of temperature-sensitive features present in the narrow-band filters. For example, the $J0410$ filter hosts a particularly 
temperature-dependent feature, the H$\delta$ line (4102\,{\AA}); see Fig.~\ref{fig:temp_HD}. In the upper left panel, we consider the strength of H$\delta$ over the temperature range 5250 $ < T_{\rm eff}$ (K) $< 8000$ for a star of log$g$ = 2.5, [Fe/H] $ = -2.50$, and [C/Fe] = 0.0. As seen in the upper right panel, this temperature range corresponds to a 31\% increase in the equivalent width of H$\delta$, from 18.4\,{\AA} to 24.2\,\AA. Thus, we anticipate success using networks in which training parameters include photometric colors and magnitudes incorporating the $J0410$ filter.

The \ion{Ca}{II} H \& K lines (3969\,{\AA} and 
3934\,{\AA}, respectively) display similar behavior with increasing metallicity. In the middle-left panel, we vary the metallicity of a $T_{\rm eff}$ = 5000\,K star over the range $-$4.00 $<$ [Fe/H] $<$ 0.0. From inspection of the middle-right panel, for a star of solar metallicity, the equivalent widths of the \ion{Ca}{II} H \& K lines are 42\% and 64\% stronger than for a [Fe/H] = $-2.0$ star of the same temperature.
However, the \ion{Ca}{II} H \& K lines also exhibit an asymmetric temperature dependence. In the same range of effective temperature, the \ion{Ca}{II} H line strength can increase by as much as 50\% for a star with [Fe/H] = $-$2.50, due to the increasing influence of H$\epsilon$ at $\lambda$ = 3970\,\AA. Meanwhile, the \ion{Ca}{II} K line nearly vanishes by $T_{\rm eff}$ = 7500\,K. We therefore expect to obtain a keen sensitivity to metallicity by incorporation of the $J0395$ filter in our methodology, and anticipate the necessity to unravel a degeneracy with effective temperature.

% #######################################
% ARTIFICIAL NEURAL NETWORK
% #######################################

\section{Artificial Neural Network}\label{network}
\subsection{Architecture}
In the development of SPHINX, we generalize the use of the core ANN element with what amounts to a three-layered system. The motivations for this structure are discussed in Section~\ref{SPHINX}. Here, we consider the basic function of an ANN.

\begin{figure}
	\centering
	\includegraphics[trim={0.3cm 1.5cm 0 0.40cm},clip, scale=0.4]{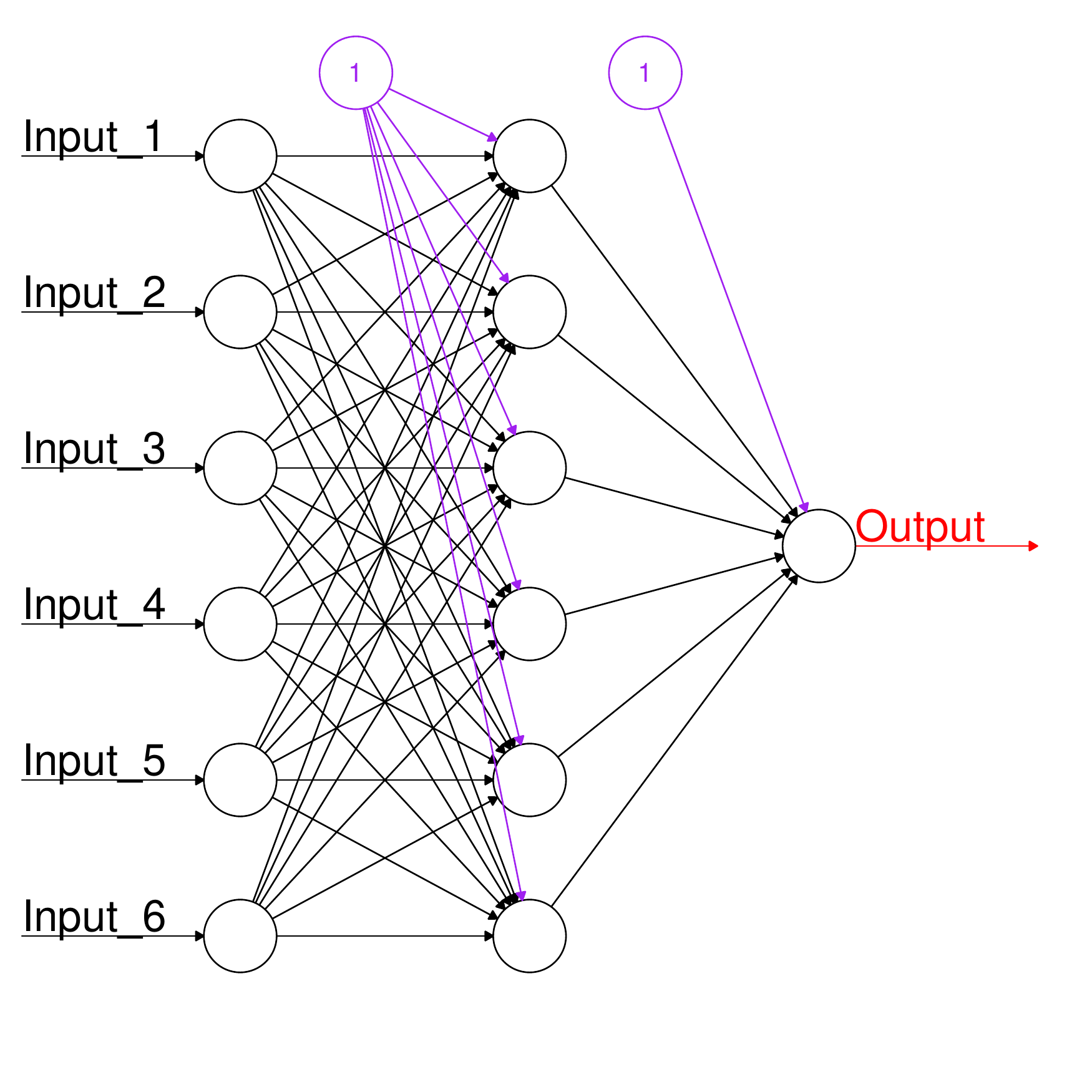}
	\caption{An example network element used in the SPHINX ANN pipeline. The base ANN utilizes a feed-forward algorithm and consists of a single hidden layer. The inputs are scaled combinations of J-PLUS colors and/or magnitudes.}
	\label{fig:net_plot}
\end{figure}

For the core implementation of the ANN, we make use of the multi-layered perceptron class, \texttt{MLPRegressor}, from \texttt{scikit-learn} \citep{Pedregosa:2011}. We choose a feed-forward algorithm with one hidden layer of neurons. As an example, Fig. \ref{fig:net_plot} shows a schematic of this network. Each node computes a non-linear function of the scalar product of the input vector and a weight vector. For \textit{N} nodes, the functional form can be expressed as follows:

\begin{equation}
y(\mathbf{x}) = \tilde{g}\Big[ \sum_{i=1}^{N}\omega_{i 2} \cdot g( \boldsymbol{\omega}_{i 1}^{\rm T} \mathbf{x})  \Big].
\end{equation}

\noindent Here, \textbf{x} denotes the input vector, in our case, the pre-scaled magnitudes from a subset of the J-PLUS filter set. Each neuron, $i$ of $N$, in the hidden layer receives the sum of the input vector with the weight vector corresponding to the first layer, $\boldsymbol{\omega_{i 1}}$, represented for simplicity by the inner product, where \textbf{T} denotes the transpose. For each hidden layer, we implement a bias neuron, the corresponding weight of which is included in $\boldsymbol{\omega_{i 1}}$. These bias neurons produce a constant output of 1, and thus are not connected to the previous layer. Bias neurons help to shape the output of the activation function, where they shift the inner product in a way analogous to the y-intercept of a linear equation.

The node then applies an activation function, \textit{g}, and the output is provided to the output layer. A similar sum is then performed with the outputs of each neuron in the hidden layer and the weights of the second layer where, in general, the output activation function, $\tilde{g}$, is not required to be the same function used in previous layers.

A hyperbolic tangent activation function was implemented in SPHINX, in part because it resulted in faster convergence times, but also because the function maps the input vectors to ($-$1,1), while the logistic sigmoid produces values in an asymmetric range (0,1). It has been empirically shown that symmetric activation functions produce faster convergence times \citep{LeCun:1991}. This is the case for the application of effective temperature and metallicity estimates.

Optimal weights for the network are determined using a stochastic gradient descent algorithm. 
With traditional back-propagation, partial derivatives are computed for each weight with respect to a specified error function, \textit{E}, to determine the influence of small perturbations of each weight's value. In this manner, new values of the $i$th weight in the $j$th layer can be determined as follows \citep{Riedmiller:1994}:

\begin{equation}
\omega_{ij}(t+1)=\omega_{ij}(t) - \epsilon \frac{\partial E}{\partial \omega_{ij}}(t).
\end{equation}

\noindent Here, $\epsilon$ is the learning rate, which scales the update correction. Fixed values of this parameter were found to introduce oscillations, where the updated value over-corrects with each iteration, and so never converges to the optimal weight. To overcome this, a stochastic gradient-based optimizer known as \textit{Adam}  was implemented, in which an adaptive-learning rate is computed from first-order moments of the gradient in the scalar error function \citep{Kingma:2014}.

%
%_____________________________________________________________
\begin{figure*}
	\centering
	\includegraphics[trim={0.3cm 3.5cm 0 0.40cm},clip, scale=0.90]{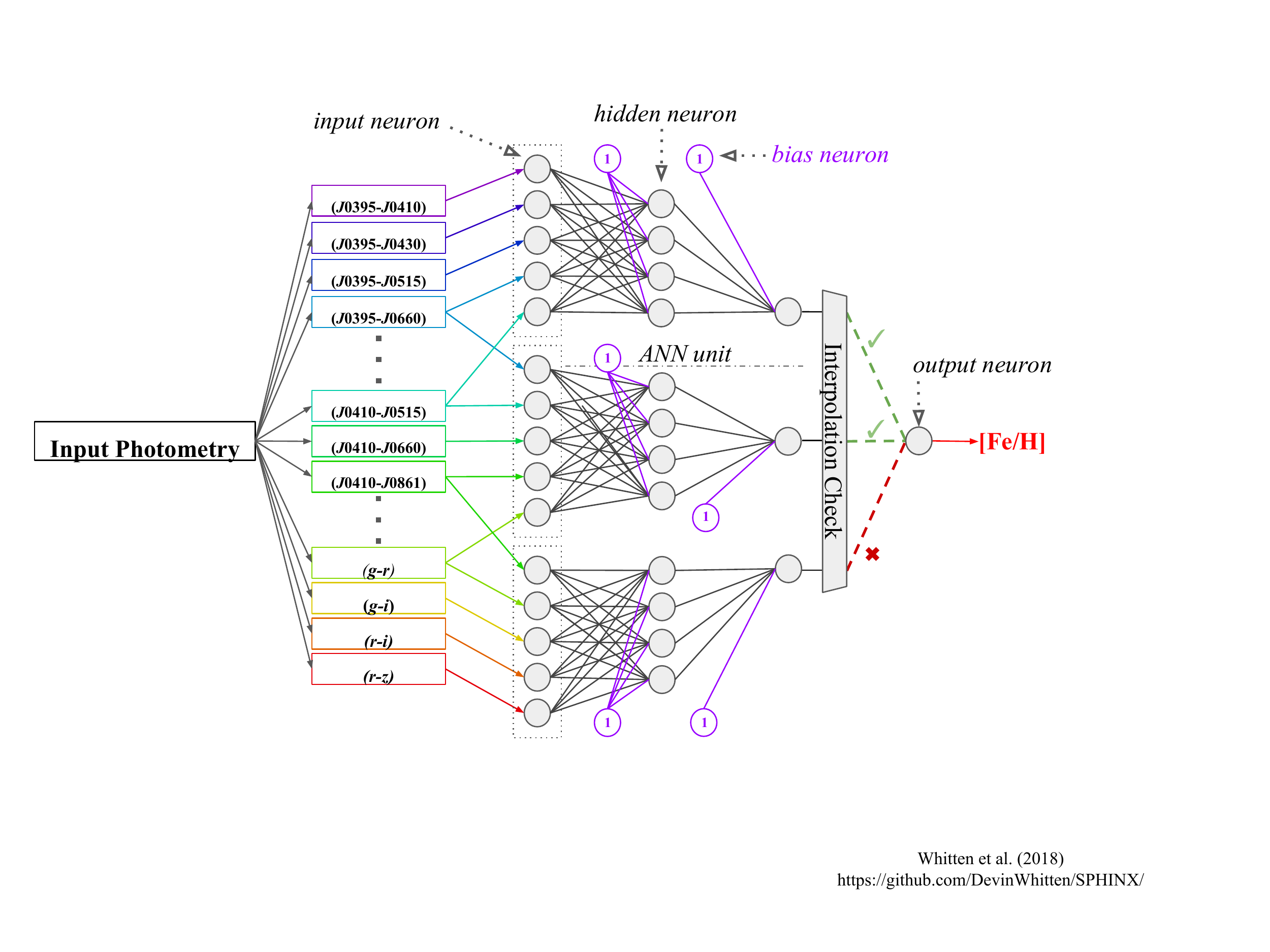}
	\caption{Example schematic of the Stellar Photometric Index Network Explorer (SPHINX). The architecture consists of a scalable three-layered neural network. The hidden layer performs the hyperbolic tangent activation function on the neuron-specific weighted inputs. The following layer represents each sub-networks best estimate of the stellar parameter. The final layer performs a validation weighted sum of the estimates verified to have resulted from the interpolating networks.}
	\label{fig:SPHINX_outline}
\end{figure*}
\subsection{Pre-scaling/Normalization}

Prior to using the magnitudes of our training set to converge the network, a form of pre-scaling must be applied, whereby all input variables are set to the same scale. The most common technique is simply a linear rescaling, in which we subtract the sample mean and divide by the standard deviation, leaving each input vector with a mean of zero and variance of one. A more robust method of pre-scaling is included as an option in SPHINX, in which the center of the distribution is taken as the median, and the spread is determined by some specified percentile range. This provides the option to control the fraction of inputs between [$-$1, 1], or the optimal domain of the activation function's response.

Although synthetic magnitudes are calibrated to the J-PLUS photometric system, it is highly desirable to train the ANN with inputs for which the distributions match that of the validation, or science set. Otherwise, pre-scaling the input colors could push these inputs to a region that requires the ANN to extrapolate from the inputs, or operate near the asymptotic region of the activation function.

This ANN structure serves as the basis for the architecture of SPHINX, where we generalize the use of the ANN in a larger, three-layered structure in an attempt to maximize the capability of the ANN, while including the variety of input combinations available. We describe the detailed design of SPHINX in Section~\ref{SPHINX_design}.

%-------------------------------------------

\section{Parameter determination and optimization with SPHINX}\label{SPHINX}

\subsection{Training set assembly}
In contrast to an ANN's capability for modeling patterns in high-dimensionality space, their ability to extrapolate beyond the boundaries of their underlying training set is quite limited. One consequence of this limitation is the need to be aware of the distribution of the validation set prior to training of the network. While we have no a-priori knowledge of the chemical abundance and atmospheric-parameter distributions in EDR and DR1, we have all of the necessary information regarding the magnitude distributions, which can be used to construct an ideal training set.

At present, SPHINX allows for selection of training stars from a variety of catalogs, including the subset of EDR and DR1 sources with available SSPP-estimated parameters, as well as SDSS Reservoir sources with synthetic magnitudes calibrated to the J-PLUS photometric system. SPHINX then performs a signal-to-noise, photometric error, and $E(B-V)$ rejection according to the specified thresholds, in addition to the faint/bright limits, all of which are set in the input parameter file. With the appropriate parameter bounds and rejections determined, the training set can be uniformly sampled across the target variable, in this case $T_{\rm eff}$ or [Fe/H]. This routine is optional, but is designed to protect against overemphasis by the network on a particular region of the parameter space. 

Once all desired processing of a training set is complete, we define a scale frame, which describes each input distribution in the target set. This includes the center location and spread estimates that SPHINX uses to linearly scale and unscale all inputs to the networks for training and parameter estimation. If the number of target sources is large, it may be desirable to set the scales according to the target photometry. However, if the target list consists of only a few sources, this is not advised; the scales should be set based on the training set's photometric distributions. The default manner of determining the center and spread of each distribution is to fit a Gaussian to each input. A more robust method utilizing the median and fourth-spread, or $f$ spread (i.e., the interquartile range), is also available, although it was found that alternative estimates of the scales did not significantly influence the network's performance.

\subsection{Approximating input distributions}

While the SDSS Reservoir and DR1 catalogs are sufficiently large that we can construct training sets across a reasonably wide range of stellar parameters and magnitudes, an effective method to accommodate a variety of target (or testing) distributions is required. For instance, our selection from the J-PLUS M15 photometry (details in Section \ref{M15_section}) comprises stars of 14 $ < g <  18$, well within the SDSS Reservoir distribution. However, the SDSS Reservoir distribution peaks at $g = 18.3$, with a standard deviation of 1.2\,mag, while the M15 distribution mean and standard deviation are $g = 16.3$ and 1.1\,mag, respectively. Training stars naively sampled from the SDSS Reservoir catalog would introduce a bias towards the fainter stars in the M15 cluster, which is not ideal. 

We include the option to force the training set input distributions to best approximate those of the target or science set. 
For a certain magnitude or color input, it is assumed that the center and spread of the target distribution have previously been set. The following error function is minimized for $\gamma$, the scale of the Gaussian distribution in the training set corresponding to the center and scale of the target distribution:

\begin{equation}
\Phi(\gamma | \bar{x}, \sigma) = \sum_{i}^{N} \left|y_i - \gamma \cdot \rm{exp}\left[\frac{-(x-\bar{x})^2}{2\sigma^2}\right]\right|^{-\alpha},
\end{equation}

\noindent where the scale $\gamma$ sets the maximum number of stars with input value $x\pm \delta x$ that can be drawn from the training distribution, such that the resulting input distribution conforms to a Gaussian described by $\bar{x}$ and $\sigma$. We leave $\alpha$ as a free parameter (default $\alpha = 1/2$) to tune the level with which the algorithm penalizes non-conformities to the desired distribution.

\subsection{Network optimization with SPHINX}\label{SPHINX_design}

In general, the proper combination of inputs for the network -- or, for that matter, their quantity -- is not known. We anticipate the use of temperature-sensitive filters such as $J0410$ and $J0660$, and metallicity-sensitive filters such as $J0395$, $J0515$, and $J0861$. Beyond these, however, it is not directly apparent what additional combinations facilitate the capabilities of the network. Further, for networks operating with different combinations of photometric inputs, one network may be forced to extrapolate, but not another. 

\par To address these concerns and ensure resilience against potentially faulty inputs in both the training and science case data, we enable a system of networks, or ANN units, each with a distinct combination of the available inputs. SPHINX determines all combinations of the specified filters and their corresponding colors, assigning and training a network for a specified number of these combinations. Throughout the operation and analysis of SPHINX, the number of inputs to each ANN unit is small ($N=6$), to maximize the ANN's dependence on the each input in the subset, and also to mitigate the convergence time of each ANN in the array. 

The basic schematic of SPHINX is shown in Fig. \ref{fig:SPHINX_outline}. For simplicity of this illustration, only a sub-sample of available colors is shown, and the number of ANN units is limited to three. As can be seen, SPHINX amounts to a three-layered neural network, consisting of a layer of subordinate ANN units that we call the network array. This array is of scalable size; we evaluate the performance of SPHINX as a function of array size in Section \ref{feh_section}. For every network estimate made, SPHINX notes when that network is extrapolating outside of its training domain. This is accomplished with a structure that we refer to as an interpolation frame, which stores the domains of all inputs in the training set. All estimates made in the network array layer are subject to an interpolation check, flagging any/all estimates where extrapolation was necessary.

During training, the performance of each ANN unit in the network array is evaluated and recorded using a validation set, and this analysis is used to assign each ANN unit a score for future reference and for computing final estimates. We define a score for the network, which we base on a modified median absolute deviation (MAD) of the verification set:
\begin{equation}
\xi_N = \rm{median}(|\mathbf{E_N} - \rm{median}(\mathbf{E_N})|)^{-\beta},
\end{equation}

\noindent where \textbf{E$_N$} represents the error in network \textbf{N}'s estimate of the target variable in the verification set. Here, $\beta$ is left as a free parameter (default $\beta=2$) to control the degree to which we penalize ANN units with poorer performance in the final parameter estimate. We then assert that the proper science estimate is the weighted sum of each network estimate and its corresponding score, $\xi_N$.

\begin{figure*}
\centering
\includegraphics[trim={1.5cm 0.1cm 0 0.40cm},clip,width=\textwidth]{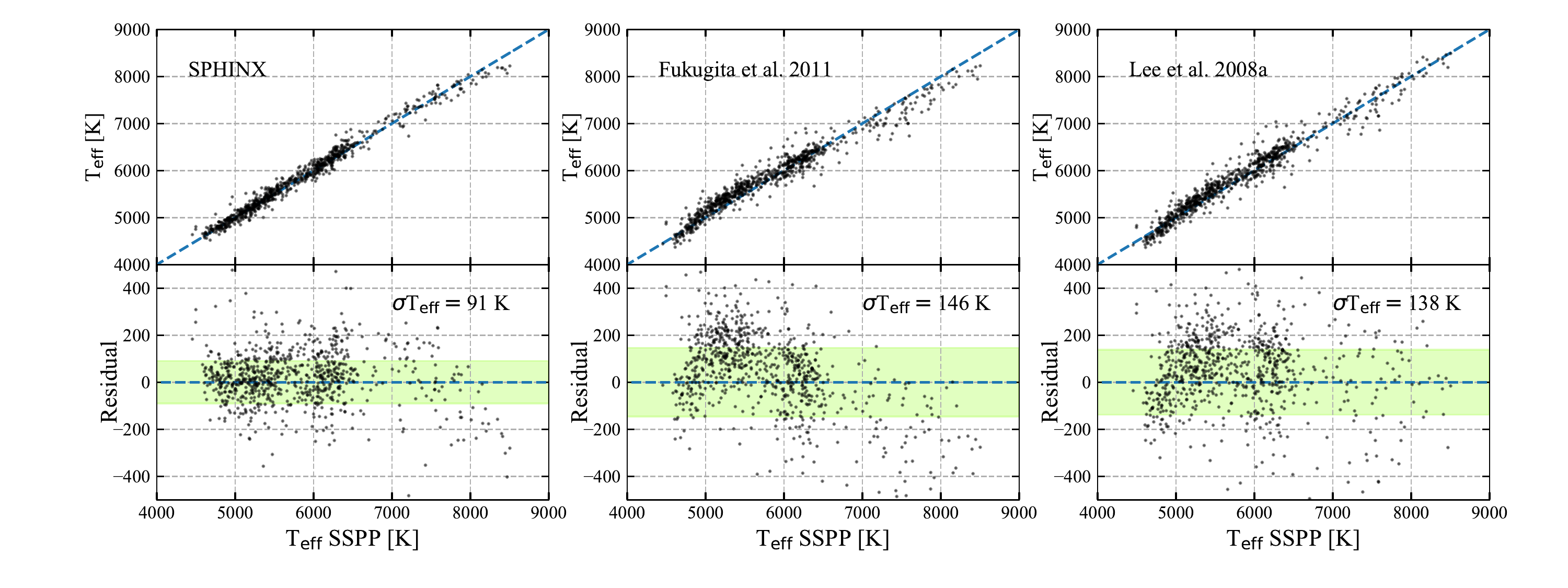}
\caption{Effective-temperature estimates, $T_{\rm eff}$, from SPHINX, as a function of the adopted value from the SSPP for a testing catalog consisting of J-PLUS EDR photometry (\textit{left panels}). For comparison, $g-r$ calibrations from \citet{Fukugita:2009} (\textit{middle panels}) and \citet{Lee:2008a} (\textit{right panels}) were applied to the sample. The residuals for each temperature calibration are shown against the accepted temperature in the bottom panel, where the green region depicts the standard deviation from a Gaussian maximum-likelihood fit.}
	
\label{fig:teff_net_result}
\end{figure*}

\par A system of twelve filters corresponds to $\binom{12}{2}$ = 66 possible colors. With a network array size of six ANN units, the resulting number of unique color combinations is $\binom{66}{6}$ = 90,858,768, an infeasible number of networks to train in a finite time.  To address this, we allow for a specified number of input combinations to be utilized, whereby a subsample of networks from the underlying ensemble is selected. This can be done at random, or if any specific filters or colors are desired, SPHINX can be restricted to employ combinations of colors in which at least one input is comprised of these specifications. We make use of this feature for effective temperature and metallicity estimates, where SPHINX is forced to use of $J0410$ and $J0395$, respectively, in combination with the other available filters.

Before consideration of the final parameter estimates, each ANN unit estimate is verified by the interpolation check described above. The number of contributing ANN units is recorded for future consideration via the \texttt{NET\_ARRAY\_FLAG}, and the final estimate is simply a weighted sum of the contributing ANN unit estimates and corresponding validation scores, $\xi_N$. With the ensemble of contributing ANN units, we compute the MAD of the estimates, as well as the standard deviation, which is also weighted by the validation scores. These serve as the uncertainty in the parameter estimate reported by SPHINX.
This procedure can also be used to determine the most optimal networks achieved in the network array. If specified, SPHINX considers only networks with the highest scores, $\xi_N$, and excludes all other estimates from the final parameter estimation.

%%%%%%%% STELLAR PARAMETER SPECIFIC METHODOLOGY HERE

\section{Effective temperature determination with SPHINX}\label{teff_section}

For effective-temperature training, we restricted the range of interest to 4500 < $T_{\rm eff}$ (K) $< 8500$. We selected stars with DR1 photometry from the SDSS Reservoir, and reserved all EDR sources for testing purposes. For simplicity, stars for which metallicities exceed Solar were excluded, anticipating the influence of high metal abundance on the underlying continua. A limit of 0.1\,mag in the error reported for the observed photometry, with a faint limit of 20 on all magnitudes was applied to the training set, in addition to the extinction limit of $E(B-V) < 0.05$\,mag stated previously. 

It was possible that, even with excellent synthetic photometric errors and an optimal $S/N$ in the underlying spectrum, the temperature assigned to the training star by the SSPP was incorrect, or at least imprecise. We therefore implemented a $\pm$ 120\,K cut on the adopted $T_{\rm eff}$ error estimate, and insisted that the individual estimates from the SSPP -- in this case the adopted estimate and estimates derived from the H$\alpha$ \& H$\delta$ Balmer-line strengths\footnote{\texttt{T1} and \texttt{T2} as described in \citet{Lee:2008a}} -- not differ from by more than 75\,K.

With the optimally cleaned stars selected, training sets were constructed by uniformly sampling temperatures between the maximum and minimum temperature thresholds. The act of sorting and partitioning the catalog into 15 bins prior to randomly sampling 200 stars from each bin ensured that a roughly even distribution of temperatures was obtained. The final DR1 training batch consisted of 1152 stars. Both the training and testing sets included stars with surface gravities in the range $1.0 < \log{g} <5.0$. The median $\log{g}$ in both sets was 4.1, with a standard deviation of 0.5\,dex. Both sets were dominated by main-sequence stars, but $\sim$5\,\% of stars had surface gravities consistant with giants ($\log{g} < 3.0$). 

SPHINX was applied to 1015 stars with EDR photometry. For analysis of the network performance, we excluded testing sources with spectral $S/N <$ 20, and/or a larger uncertainty in the SSPP $T_{\rm eff}$ estimate ($\sigma(T_{\rm eff}) > 150$\,K). For comparison, effective temperature estimates were made with two broad-band calibrations, Eq. 1 of \citet{Fukugita:2009}, and $T_l$ of \citet{Lee:2008a}. \citet{Fukugita:2009} determined an empirical temperature range for their calibration of $3850 < T_{\rm eff}$ (K) $< 8000$, and noted the smallest scatter of $\sigma(T_{\rm eff})=93$\,K when using $g-r$ compared to other broad-band colors. The $T_l$ estimate was derived to be effective for a wide range of temperatures, beyond $4500 < T_{\rm eff}$ (K) $< 7500$, making it ideal as a comparison for our calibration with SPHINX.

We made use of a number of robust estimators of central location and scale, and adopted the suggested nomenclature from \citet{Beers:1990}. For measures of central location, in addition to the median $C_{\rm M}$, we used the trimean $C_{\rm TRI}$ -- based on the sample median and upper and lower fourths of the distribution(s) -- the biweight location $C_{\rm BI}$, and the mean $C_{\mu}$. For estimates of scale, we made use of a normalized median absolute deviation, $S_{\rm MAD}$, which is particularly suited for heavy-tailed distributions \citep{Beers:1990}. In addition, a scale estimate based on the fourth spread, the $f$ pseudosigma, $S_{f}$, was used, along with the biweight scale estimate $S_{\rm BI}$ and standard deviation $S_{\sigma}$. 

The results of the temperature estimates made with SPHINX and the comparison calibrations on the EDR photometry are shown in Fig. \ref{fig:teff_net_result} and summarized in Table \ref{table:temp_stats}, where we compare as well the statistics of high temperature stars ($T_{\rm eff} > 7000$\,K). For estimates made with SPHINX, shown in the leftmost panel of Fig. \ref{fig:teff_net_result}, central location estimates of the residuals $C_{\mu}=+21$\,K and $C_{\rm TRI}=+20$\,K indicated a slight bias towards over-estimation for the EDR sample. This behavior was in contrast to the under-estimation seen for temperatures above $T_{\rm eff} > 7000$\,K, where the $C_{\rm M}$ and $C_{\rm TRI}$ of the residuals were $-14$\,K and $-28$\,K, respectively. The scatter in the region, where $S_f$ increased from 93\,K to 171\,K. Similarly, the $S_{\rm MAD}$ increased from 93\,K to 172\,K. We attributed this behavior both to edge effects emerging near the limit of the network interpolation range. Interestingly, it was found that inclusion of the $J0410$ filter did not significantly improve the performance of effective temperature estimates made with SPHINX. We conclude that the quality of these estimates is more likely a matter of J-PLUS filters tracing the overall SED, as well as the capability of SPHINX to interpret the behavior of the SED over a wide temperature range.

\begin{table}
	\caption{Central location and scale estimates of photometric temperature calibration residuals.}             % title of Table
	\label{table:temp_stats}      % is used to refer this table in the text
	\centering                          % used for centering table
	\tiny
	\begin{tabular}{c | c c c}        % centered columns (4 columns)
		\hline\hline                  % inserts double horizontal lines
		& SPHINX & Fukugita et al. 2011 & Lee et al. 2008a\\  % table heading 
		
		& (K) & (K) & (K) \\
		\hline   
        
        $C_{\rm M}$ & $+20$ & $+60$ & $+57$ \\
        $C_{\rm TRI}$ & $+20$ & $+56$ & $+56$ \\
        $C_{\rm BI}$ & $+20$ & $+57$ & $+56$ \\
        $C_{\mu}$ & $+21$ & $+66$ & $+64$ \\
        \hline
        $S_{\rm MAD}$ & 93 & 157 & 143\\
        $S_{f}$ & 93 & 156 & 144\\
        $S_{\rm BI}$ & 106 & 171 & 156\\
        $S_{\sigma}$ & 91 & 146 & 138\\
        \hline
        \multicolumn{4}{c}{$T_{\rm eff} > 7000$\,K} \\
        \hline
        $C_{\rm M}$ & $-14$ & $-230$ & $-10$ \\
        $C_{\rm TRI}$ & $-28$ & $-233$ & $-15$ \\
        $C_{\rm BI}$ & $-26$ & $-230$ & $-19$ \\
        $C_{\mu}$ & $-5$ & $-201$ & $+13$\\
        \hline        
        $S_{\rm MAD}$ & 172 & 218 & 199 \\
        $S_{f}$ & 171 & 213 & 195 \\
        $S_{\rm BI}$ & 161 & 195 & 208 \\
        $S_{\sigma}$ & 188 & 196 & 196 \\
		% inserts single horizontal line
		
		\hline                                   %inserts single line
	\end{tabular}
\end{table}

Using the $g-r$ calibration given in Eq. 1 of \citet{Fukugita:2009}, we computed an $S_{f}$ of 156\,K. This calibration, shown in the middle panels, was biased as well towards over-estimation, indicated by the mean of the residuals, $+$66\,K, and $C_{\rm TRI}$ of $+$55\,K. Similar to the estimates made with SPHINX, we found under-estimation of effective temperature beyond $T_{\rm eff} > 7000$\,K, where the median of the residuals was $-$230\,K, with a $C_{\rm TRI}$ of $-$233\,K. Estimates in this regime became increasingly uncertain, with $S_f$ of 213\,K, and a $S_{\rm MAD}$ of 218\,K.

In the right panel we compare the network temperature estimates to a polynomial calibration of $g-r$ from the SSPP, $T_I$ \citep[see][for details]{Lee:2008a}. For this calibration, the median in the residuals of $+$57\,K and $C_{\rm TRI}$ of $+$56\,K again indicated a systematic over-estimation. The standard deviation of the Gaussian maximum-likelihood fit of 138\,K increased to 196\,K in the high temperature range, $T_{\rm eff} > 7000$\,K. A number of outliers were apparent around $T_{\rm eff} \sim 6200$\,K for each calibration, particularly in the \citet{Lee:2008a} calibration, which was expected if variable stars such as RR Lyrae stars are present in the EDR sample. Offsets such as these would occur for stars of which spectroscopic and photometric temperature estimates were made at disparate phases of the pulsation cycle.  

We concluded that the use of SPHINX with J-PLUS photometry provides effective temperatures with a bias reduced by a factor of three and a dispersion reduced by $\sim$40\% with respect to the previous broad-band calibrations. In, addition SPHINX is less prone to under-estimation in the high-temperature regime beyond 6500\,K. No significant influence by surface gravity on the effective temperature was seen in the range $2.0 < \log{g} < 5.0$, however for lower values ($\log{g} < 2.0$) determinations tended to be over-estimated by $\sim70$\,K.

\begin{figure*}
	\centering
	\includegraphics[trim={1.2cm 1.25cm 2.5cm 0.40cm},clip, width=\textwidth]{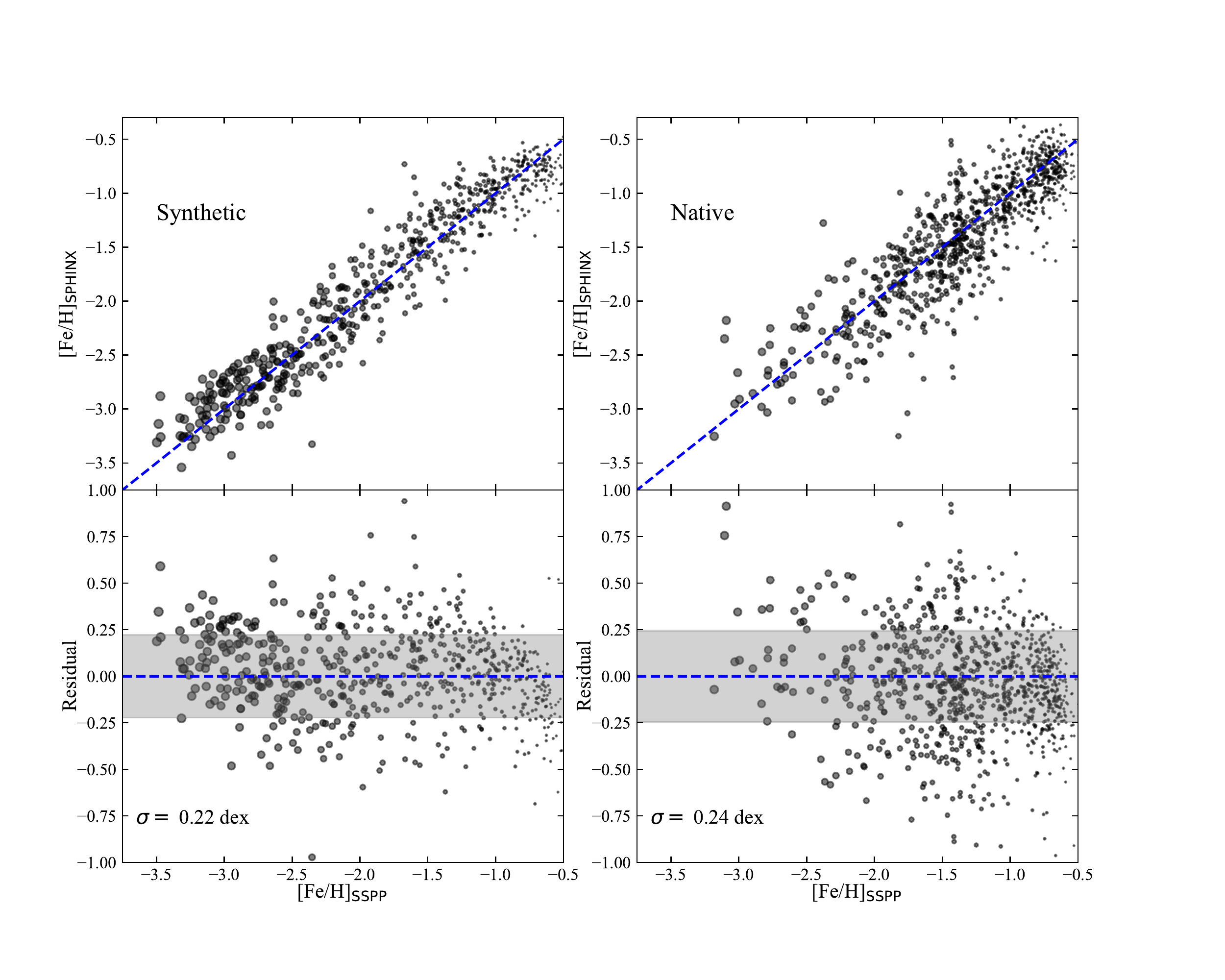}
	\caption{Metallicity estimates, [Fe/H], from SPHINX, as a function of the biweight estimate from the SSPP, for stars of effective temperature $4500<$ $T_{\rm eff}$ (K) $<$ 6200. The gray region corresponds to the standard deviation of the residuals, determined from a maximum-likelihood Gaussian fit. Results from SPHINX using synthetic magnitudes from the SDSS Reservoir (\textit{left panels}) are shown with the results from native J-PLUS DR1 photometry (\textit{right panels}) for comparison.}
	\label{fig:feh_net_result}
\end{figure*}

\begin{table*}
	\caption{Central location and scale estimates of photometric metallicity determinations.}           % title of Table
	\label{table:feh_stats}      
	\centering                          
	\begin{tabular}{c | c c c c | c c c c}      
		\hline\hline                 
		Trial & $C_{\mu}$ & $C_{\rm M}$ & $C_{\rm TRI}$ & $C_{\rm BI}$ & $S_{\sigma}$ & $S_{\rm MAD}$  & $S_{f}$ & $S_{\rm BI}$ \\  % table heading 
		& (dex) & (dex) & (dex) & (dex) & (dex) & (dex) & (dex) & (dex)\\
		\hline   
        %&&&&&&&&\\
        SDSS Reservoir & +0.02 & +0.03 & $+0.03$ & $+0.02$ & 0.22 & 0.22 & 0.22 & 0.23\\
        J-PLUS DR1 & $-0.06$ & $-0.06$ & $-0.05$ & $-0.06$ & 0.25 & 0.25 & 0.25 & 0.27 \\
		\hline                                   %inserts single line
	\end{tabular}
\end{table*}

\section{Metallicity determination with SPHINX}\label{feh_section}

Two separate trials were conducted for validation of the metallicity routine in SPHINX. The maximum error on the biweight estimate of [Fe/H] from the SSPP was set to $\pm 0.20$\,dex. Similar to the temperature training, we insisted that the adopted and biweight estimates of metallicity from the SSPP did not differ by more than $\pm$0.20\,dex.
We refer the interested reader to \citet{Lee:2008a} for an in-depth description of the biweight and adopted estimators from the SSPP. In short, the biweight is simply the robust average of all accepted estimates from the pipeline, while the adopted estimate takes into account an average of the biweight and a refined estimate that considers the reduced $\chi^2$ of a synthetic spectrum match. In the event that the refined and biweight estimates do not differ significantly ($<$ 0.15\,dex), the adopted value is set to the biweight estimate. The reliability of the metallicity estimates from the SSPP was somewhat dependent on the $S/N$. We therefore set the minimum $S/N$ to 40. The faint limit for all magnitudes was set to 18.5 to insure reliability in the synthetic magnitude calibrations.

For both trials, an array of 100 ANN units was constructed, where SPHINX was set to consider the five highest performing networks in each array. Anticipating the utility of the $J0395$ filter sensitivity to the \ion{Ca}{II} H \& K feature, we insisted on the use of $J0395$ photometry in each of the ANN units. The results of the metallicity estimates for both samples are shown in Fig.~\ref{fig:feh_net_result}. Estimates of spread and central location in the residuals are provided in Table.~\ref{table:feh_stats}.

\subsection{J-PLUS DR1 trial}
\par The first trial implemented a testing set consisting entirely of J-PLUS DR1 photometry for stars in the range of $4500<$ $T_{\rm eff}$ (K) $<6200$. The subset of the SDSS Reservoir with DR1 photometry in this range, after the error criteria described previously, was limited to 935 stars, which included only 28 stars with [Fe/H]~$< -2.5$, of which five had [Fe/H]$<-3.0$. Therefore, all DR1 stars with [Fe/H]~$<-2.5$ were reserved for the testing set, and we supplemented the DR1 training photometry with 500 stars with [Fe/H]~$<-2.5$ and synthetic magnitudes from the SDSS Reservoir, of which 250 had [Fe/H]~$<-3.0$. By doing so, the testing set consisted entirely of native J-PLUS photometry, while the training set maximized the number of low-metallicity stars in the sample. The surface gravity range for the testing set was $1.3 < \log{g} < 4.8$, with a median of 4.2 and standard deviation of 0.7\,dex.
\par The median, $C_{\rm M}$, and biweight central location, $C_{\rm BI}$, of the residuals were both found to be $-0.06$\,dex for the DR1 trial, so the predictions were prone to slight under-estimation when compared to the SSPP values. The Gaussian fit to the residuals revealed a standard deviation of $S_{\sigma}=0.25$\,dex. The largest estimate of scale for the DR1 trial was the biweight estimate, $S_{\rm BI} =0.27$\,dex. The scatter in the metallicity was independent of surface gravity in the range of $1.75 < \log{g} < 4.5$. SPHINX tended to over-estimate the metallicity for low surface gravity stars ($\log{g}<1.75$), for which the residual median increased to $C_{\rm M} = +0.14$\,dex. However, these stars are correspondingly cool ($T_{\rm eff}\lesssim4750$\,K), and it was found that stars in the range $4500 < T_{\rm eff}$ (K) $<4850$ tended to produce over-estimations regardless of the surface gravity.

\par For both the classification and recovery fractions of VMP stars with SPHINX, errors were expressed using the Wilson score approximation of the binomial confidence interval \citep{Wilson:1927,Brown:2001}. Of the 81 VMP stars in the DR1 sample, $85 \pm 3$\,\% were recovered by the [Fe/H] estimate from SPHINX. SPHINX classified 109 stars in the DR1 sample as VMP, in which $63 \pm 4$\,\% were confirmed by the biweight estimate from the SSPP.

\subsection{Synthetic SDSS trial}

\par To explore the extent of current metallicity sensitivity with SPHINX, the second trial consisted of training and testing sets constructed from the SDSS Reservoir, using synthetic magnitudes. Effective temperature for both training and testing sets was again restricted to $4500<$ $T_{\rm eff}$ (K) $<6200$, with a surface gravity range of $1.0 < \log{g} < 4.8$. We included a wider distribution of surface gravities in the synthetic SDSS trial, with a median $\log{g}$ of 3.8 and standard deviation of 0.78\,dex, to further explore the potential influence on the final determinations. The training set consisted of 1986 stars, 590 of which had [Fe/H] $<-2.5$, with 269 having [Fe/H]~$<-3.0$. 

\par A slight over-estimation was seen in the residuals of the synthetic SDSS trial, where the median and trimean estimates were both $+0.03$\,dex. Sensitivity was found to diminish below [Fe/H]$<-3.0$, where the residual median, $C_{\rm M}$, increased to 0.17\,dex. The standard deviation of the residuals was $S_{\sigma}=0.22$\,dex, somewhat improved from the DR1 trial. As seen in Table~\ref{table:feh_stats}, all estimates of scale were found to be smaller for the synthetic SDSS trial. Similar to the DR1 trial, scatter in the residuals was independent of surface gravity in the range $1.5 < \log{g} < 5.0$. Over-estimation was also seen for cooler stars in SDSS trial, where the median residual increased to $C_{\rm M}=+0.32$\,dex for stars of $T_{\rm eff} < 4750$\,K. The synthetic SDSS trial was repeated while excluding the use of the $J0395$ filter, resulting in a significant increase in scatter, where $S_{\sigma}=0.41$\,dex and $S_{\rm MAD}=0.43$\,dex. We concluded that the $J0395$ photometry is indeed a crucial component for metallicity sensitivity with SPHINX.

\par Of the 214 VMP stars in the synthetic SDSS trial, $91 \pm 2$\,\% were recovered, while $93 \pm 2$\,\% of the 209 stars classified as VMP by SPHINX were confirmed the SSPP estimate. While there is an insufficient number of EMP stars in the DR1 sample at present, we can investigate the classification and recovery fractions of EMP using the synthetic SDSS sample. Of the 43 stars determined to be [Fe/H] $< -3.0$ by the SSPP, SPHINX recovered $53 \pm 6$\,\%. Of the 33 stars classified as EMP by SPHINX, $70 \pm 6$\,\% were confirmed by the SSPP estimate, while all remaining stars were VMP. These results are comparable to those obtained by the Pristine survey \citep{Starkenburg:2017}.

\begin{figure}
	\centering
	\includegraphics[trim={0.3cm 0.25cm 0 1.50cm},clip, scale=0.78]{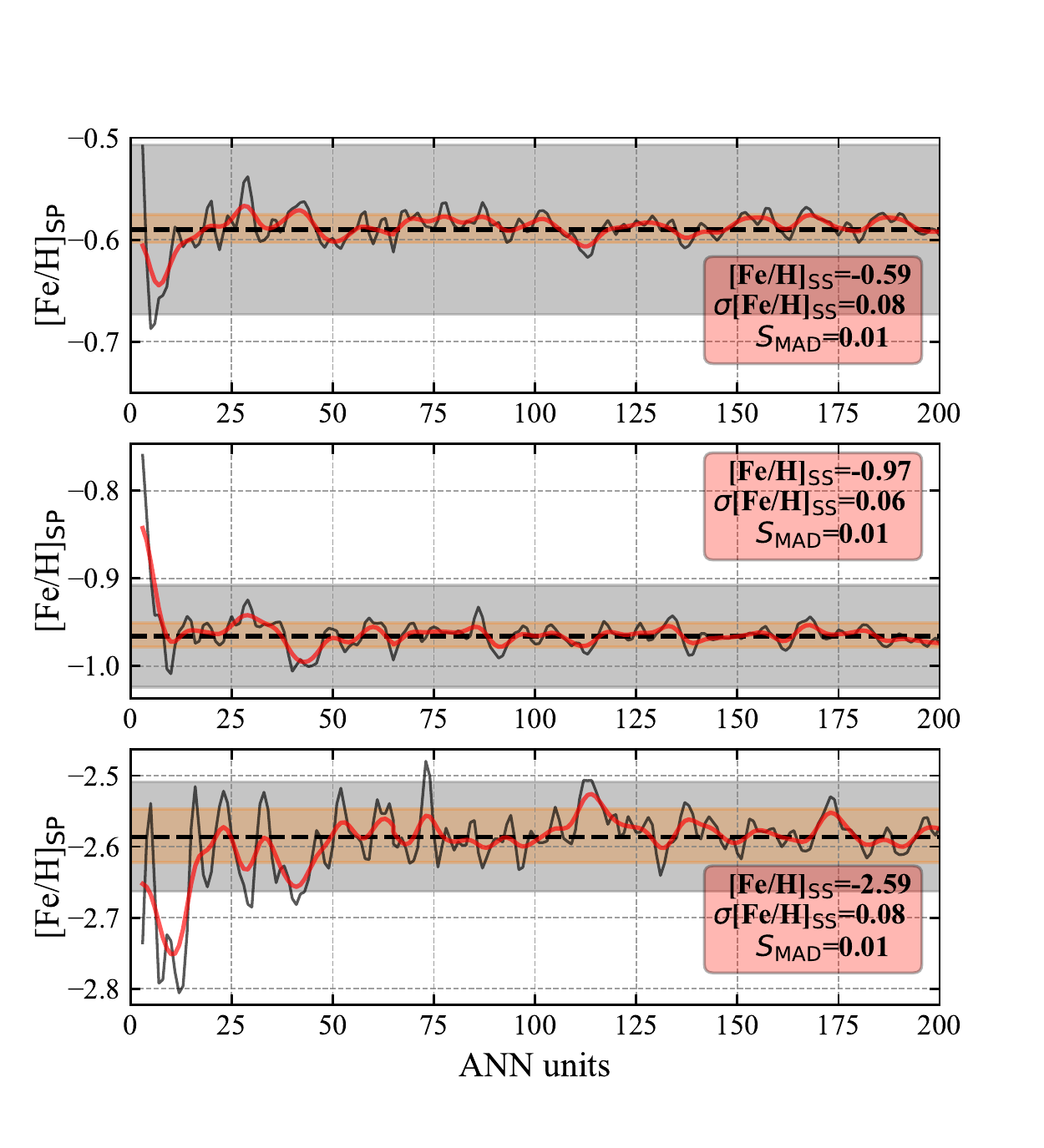}
	\caption{Metallicity estimates from SPHINX for three example stars with DR1 photometry, as a function of the number of ANN units employed. The scale estimate, $S_{\rm MAD}$, is represented by the orange band. In all three cases the scale is within the uncertainty in the spectroscopic value determined by the SSPP, shown in gray.}
	\label{fig:feh_net_array}
\end{figure}

\begin{figure*}

	\includegraphics[trim={0.5cm 0.25cm 0 0.00cm},clip, scale=0.65]{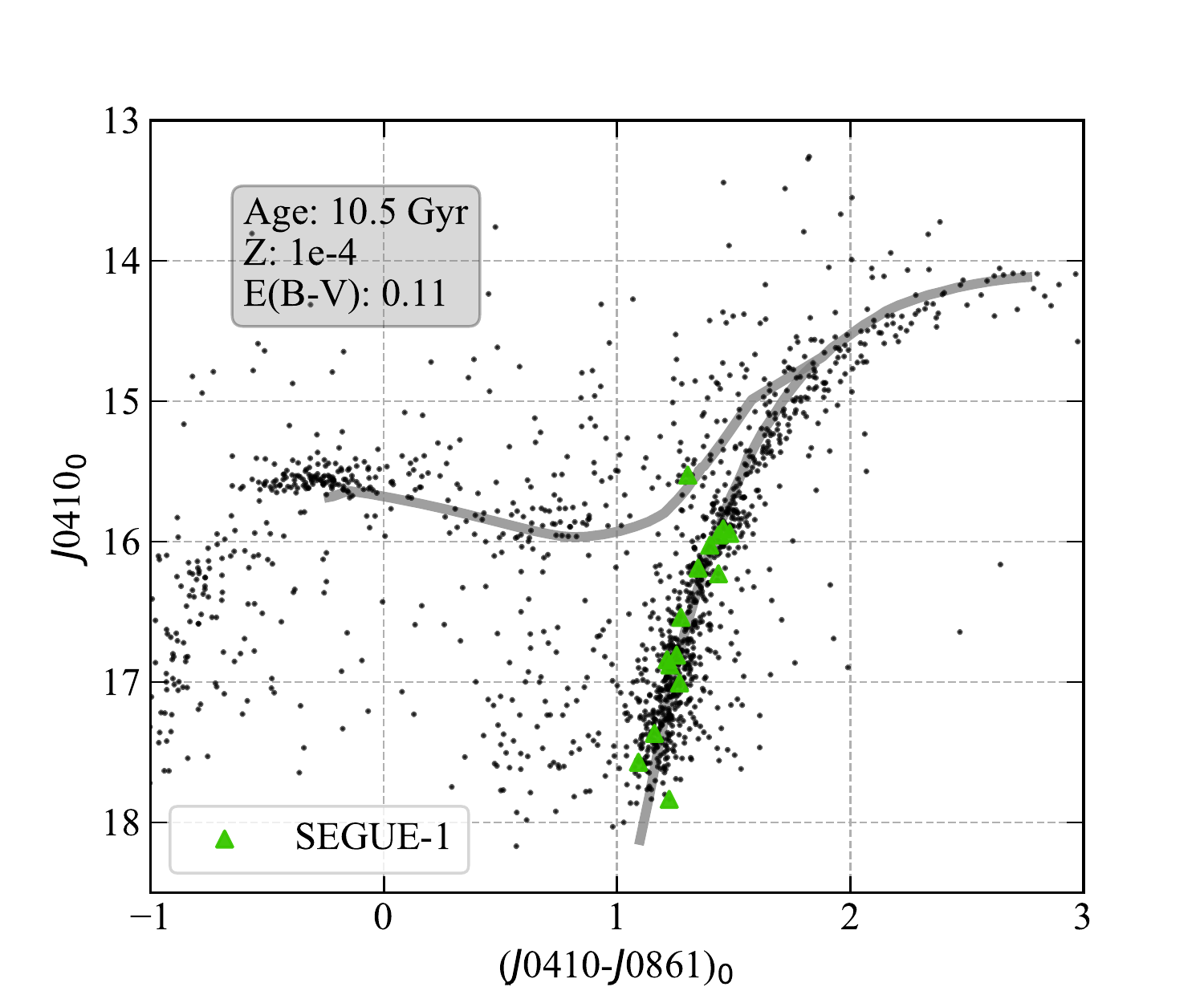}
    	\includegraphics[trim={0.00cm 0.15cm 0 0.00cm},clip, scale=0.755]{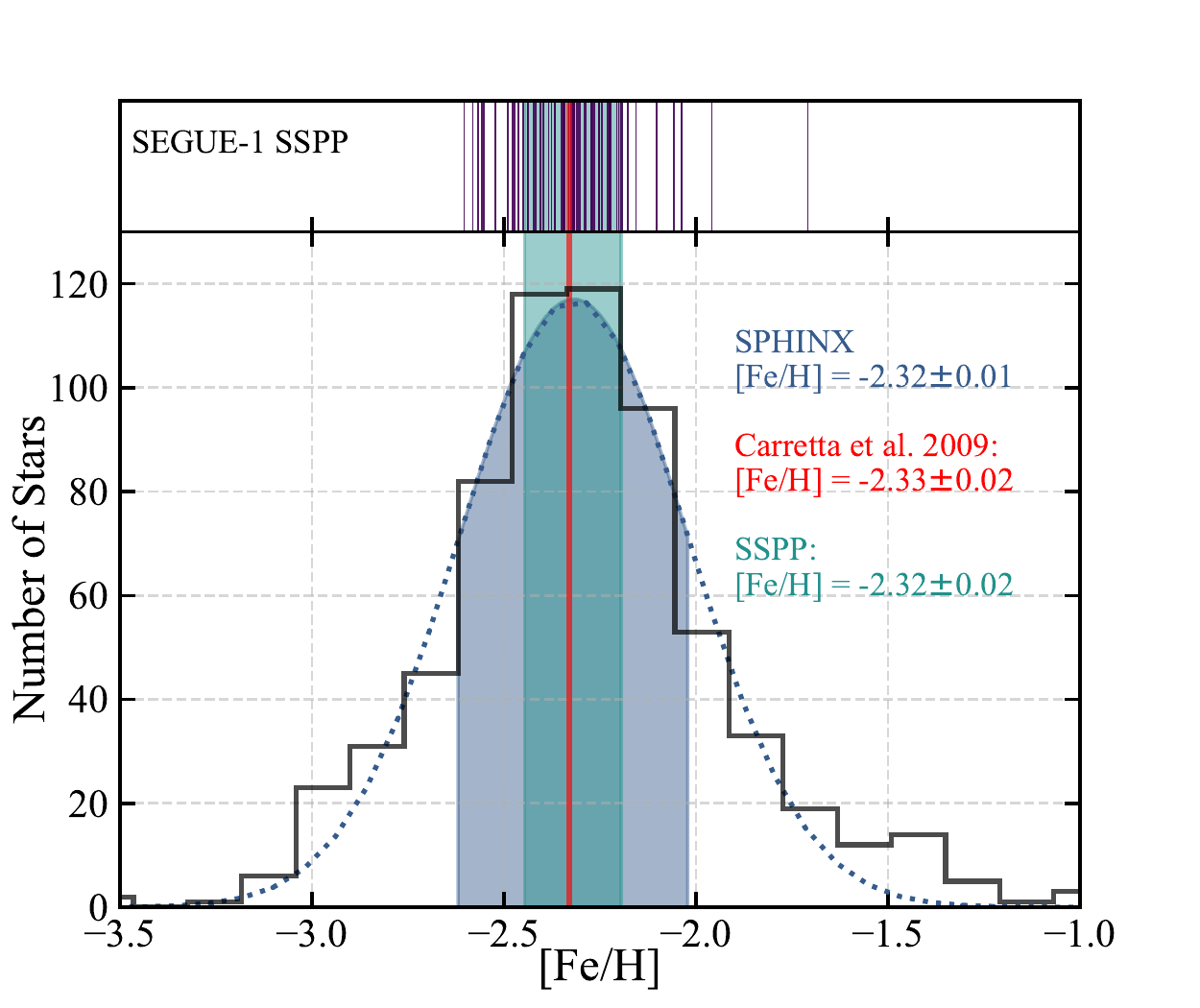}
	\caption{\textit{Left panel:} Color-magnitude diagram of the J-PLUS M15 stars. The population is over-plotted with a PARSEC isochrone corresponding to an age of 10.5\,Gyr, and metallicity of $Z=0.0001$ \citep{Bressan:2012}, with an assumed extinction of $E(B-V)=0.11$\,mag. Of the 1437 stars in the sample, the 15 with SEGUE-1 medium-resolution spectra are denoted by green triangles. \textit{Right panel:} Distribution of photometric metallicity estimates from M15. A Gaussian distribution was fit to the photometric distribution, with the $\sigma$=0.33\,dex region shaded. A stripe density plot is shown for the 66 confirmed M15 members with medium-resolution parameters from the SSPP. The shaded region in the density plot depicts the $S_{f}$ of 0.13\,dex.}
	\label{fig:isochrone_distro}
\end{figure*}

To evaluate the performance of SPHINX as a function of the number of ANN units employed in the network array, another series of trials were run on the DR1 photometric set. We made use of the DR1 training catalog supplemented with synthetic VMP magnitudes, as described above. The number of ANN units was varied from three to 500, and the classification and recovery fractions for all cases was evaluated.
We noted a small but consistent increase in the classification fraction, from 49\% to 51\% over the range of three to 100 ANN units.
We found large oscillations in the recovery fractions over the full range of employed ANN units, but a general increase of a few percent.
In the case of individual estimates, we found generally better convergence to a limiting value as the array size was increased. In many cases, estimates varied over 0.20\,dex with smaller array size. This behavior is shown for three example stars with different metallicities in Fig. \ref{fig:feh_net_array}. In all cases, the variation in the estimates was seen to stabilized after array sizes $N > $ 50. For all three cases considered, the variation in the estimates of the photometric metallicities, determined from the MAD, was well within the uncertainty reported from the SSPP estimate.

%
%_____________________________________________________________
\section{A case study with M15}\label{M15_section}

Photometric sources from the globular cluster M15 (\object{Messier~15}, NGC 7078) provide an opportunity to test both the accuracy and precision of metallicity estimates from SPHINX. Located at $\sim$10.4\,kpc from the Sun, M15 is a particularly bright globular cluster with a horizontal-branch magnitude of $V=15.8$ and tidal radius of 21.5\,arcmin \citep{Harris:1996}. M15 has a well-defined spectroscopically determined age of 10.56 $\pm$ 0.47\,Gyr \citep{Koleva:2008} and metallicity of [Fe/H] = $-2.33\pm0.02$ \citep{Harris:1996,Carretta:2009}. Furthermore, with an intrinsic scatter in the metallicity of $\sigma$([Fe/H])$< 0.05$\,dex, we regard this cluster as essentially mono-metallic \citep{Carretta:2009}. The accuracy of SPHINX determinations is thereby investigated by considering the central location of the photometric metallicity estimates for the cluster, while the spread is reflective of the precision. 

For this analysis, we made use of a stellar sample from \citet{Bonatto:2018}. Data were obtained as a science verification set prior to the J-PLUS Early Data Release. The 1.4 $\times$ 1.4\,deg$^2$ field of the J-PLUS T80Cam was sufficiently large to contain the projected area of M15 in a single pointing. Photometry extended to a limiting magnitude of $g\sim 21.5$ for over 40,000 stars. The photometric uncertainty at $g\sim21.5$ was estimated to be $\sigma_{m}\sim0.2$, however typical uncertainties for magnitudes within the calibration limit for SPHINX ($g\sim18.79$) were less than 0.05\,mag. Contrary to the standard procedure of \texttt{SExtractor} used by OAJ, \texttt{DAOPHOT PSF} \citep{Stetson:1987} was used to extract photometry of stars for M15, as the standard pipeline is not currently able to cope with the level of stellar crowding typical of GCs, and \texttt{DAOPHOT PSF} can typically reach deeper photometric limits than \texttt{SExtractor} \citep{Bonatto:2018}. \citet{Bonatto:2018} applied a decontamination algorithm using color-magnitude diagrams, based on the work of \citet{Bonatto:2007}, resulting in 1437 candidate member stars.

Our analysis included 98 stars with medium-resolution spectra from SEGUE-1. Stellar parameters and radial velocities for these stars were determined from the SSPP. For stars with high-quality estimates from the adopted and biweight procedures, effective temperatures and metallicities were taken to be the average of the two. Otherwise, a star was assigned the best of the two estimates, evaluated on the basis of a visual comparison with model synthetic spectra. Typical uncertainties in the radial velocity estimates were $\pm$7.0 km s$^{-1}$. These stars were originally selected as candidates of M15 on the basis of their proximity to the center of the cluster and thus required a check for foreground/background contamination.

We rejected non-member SEGUE-1 stars on the basis of radial velocity. First, we computed the biweight location and scale for the distribution of radial velocities in the sample. With the tuning constant of the biweight estimator set at 6.0, the central location was found to be $C_{\rm BI} = -108.0$\,km s$^{-1}$, commensurate with a $C_{\rm TRI}$ = $-107.9$\,km s$^{-1}$. For estimates of scale, we computed $S_{\rm BI}$ = 11.5\,km s$^{-1}$, $S_{f}$ = 11.5\,km s$^{-1}$.
%, and 9.9\,km$\cdot$s$^{-1}$. 
We rejected as outliers stars with radial velocities outside of the range $C_{\rm BI} \pm 2~S_{\rm BI}$. This process was repeated until no new stars were rejected; 66 stars remained. We recomputed the $C_{\rm BI}$ and $S_{\rm BI}$ again, and found $-108.0$\,km s$^{-1}$ and 6.9\,km s$^{-1}$, respectively. Finally, we compared these results with those determined by \citet{Pryor:1993}, on the basis of high-dispersion spectroscopy. They measured a mean velocity of $-107.09 \pm 0.82$\,km s$^{-1}$, and a velocity dispersion of $8.95 \pm 0.59$\,km s$^{-1}$, very similar to our values.

Of the 66 SEGUE-1 stars selected on the basis of radial velocity, a cross-match with the J-PLUS M15 catalog using a 3\,\arcsec search radius identified 15 stars in common. These stars are shown in the left panel of Fig. \ref{fig:isochrone_distro}, along with 1083 J-PLUS stars and a  $Z$=$10^{-4}$, 10.5\,Gyr isochrone generated with the Padova and Trieste Stellar Evolution Code (PARSEC; \citealt{Bressan:2012}). Here, the distance modulus for the cluster was taken to be 15.4\,mag \citep{Harris:1996}. A correction of 0.3\,mag was applied to the $(J0410-J0861)_0$ color prediction from PARSEC to align the isochrone with the giant branch seen in the M15 photometry. These spectroscopic targets largely occupied the giant branch, with one possible horizontal-branch star.

SPHINX was applied to J-PLUS photometry of the 1437 M15 stars with a limiting magnitude of $g\sim18.0$ using an array of 50 ANN units, and resulted in parameter estimates for 1041 sources (96\% of the initial sample). The remaining 396 were rejected by the pipeline due to restrictions on the photometric errors, or extrapolation of all contributing networks. 
We excluded estimates produced with less than 25 of the available ANN units, after which 664 (61\% of the initial sample) stars remained. The right panel in Fig.~\ref{fig:isochrone_distro} compares the distribution of photometric [Fe/H] estimates to both the 66 SSPP medium-resolution parameters, as well as to the external spectroscopic value of [Fe/H]$ = -2.33 \pm 0.02$ for the cluster \citep{Carretta:2009}. A central metallicity of [Fe/H] $ = -2.32 \pm 0.01$ was found from a Gaussian maximum-likelihood fit to the distribution of metallicities from SPHINX. The error estimate of the central metallicity, $\pm 0.01$\,dex, was determined from a bootstrap procedure, for which the 664 stars were randomly sampled with replacement for 1000 trials. The standard deviation of the residuals for the photometric estimates was determined to be 0.29\,dex. We compared this to the median value of [Fe/H] $ = -2.31\pm 0.02$ for the 66 SEGUE-1 stars with SSPP parameters, for which the error in the median was determined from the bootstrap procedure. The $f$ spread of the SSPP metallicity distribution was 0.17\,dex, corresponding to $S_{f} = 0.13$\,dex.

\begin{table*}
	\caption{Spectroscopic stellar parameters of 13 SEGUE-1 + J-PLUS stars in M15.}            % title of Table
	\label{table:M15}      % is used to refer this table in the text
	\centering                          % used for centering table
	\tiny
	\begin{tabular}{c c c c c c c c c c}        % centered columns (4 columns)
		\hline\hline                  % inserts double horizontal lines
		RA & DEC & $T_{\rm SP}$ & $\sigma(T_{\rm SP})$ & $T_{\rm SS}$ & $\sigma(T_{\rm SS})$ & [Fe/H]$_{\rm SP}$ & $\sigma$([Fe/H]$_{\rm SP})$ & [Fe/H]$_{\rm SS}$ & $\sigma(
$[Fe/H]$_{\rm SS}$)\\  % table heading 
		
		(hh:mm:ss) & (dd:mm:ss) & (K) & (K) & (K) & (K) &  & (dex) &  & (dex) \\ 
		\hline   
		% inserts single horizontal line
		\input{M15_latex_table.dat}
		\hline                                   %inserts single line
	\end{tabular}
\end{table*}

\begin{figure}
	\centering
	\includegraphics[trim={3.5cm 6.40cm 0 6.20cm},clip, scale=0.50]{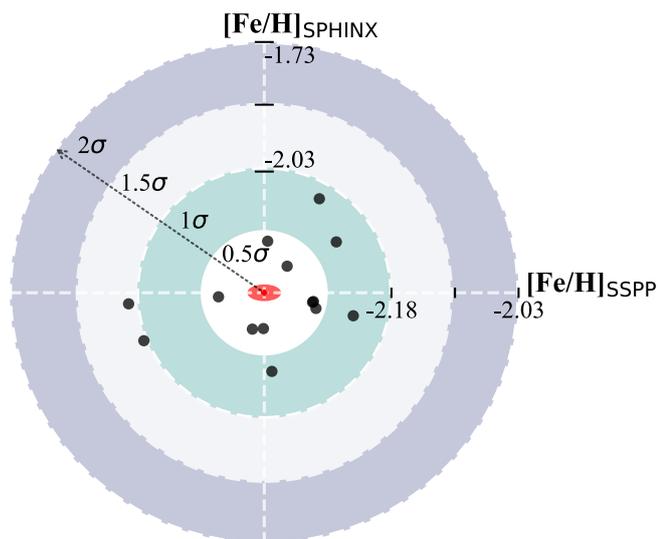}
	\caption{Metallicity estimates for 13 stars in the M15 cluster with both SEGUE-1 spectra and J-PLUS photometry. Estimates are shown above the 1$\sigma$ and 2$\sigma$ regions corresponding to the standard deviation of SPHINX estimates for the entire cluster, and the median error estimate from the SSPP. The small red ellipse in the center represents the intrinsic scatter of the M15 cluster, [Fe/H] = $-2.33 \pm$ 0.02 \citep{Carretta:2009}.}
	\label{fig:M15_one_one}
\end{figure}

We consider the accuracy of the photometric metallicity estimates for the cluster in Fig. \ref{fig:M15_one_one}. With the $\sigma$ estimate from the Gaussian distribution of 0.30\,dex and the uncertainty from the SSPP of 0.13\,dex, we depict the individual estimates for 13 stars of the original 15 with both photometric and spectroscopic estimates. The elliptical regions of Fig.~\ref{fig:M15_one_one} corresponding to 0.5, 1.0, 1.5, and 2$\sigma$ from the center value of [Fe/H] $= -2.33$ for the cluster were vertically scaled to a circular region for simplicity. We excluded one star from this sample due to a large error estimate from SPHINX ([Fe/H]$_{\rm SP} = -2.94 \pm 0.79$). Another star, thought originally to occupy the horizontal branch,  was found to have a large deviation between the photometric and spectroscopic estimates. We therefore excluded this estimate on the basis of potential error in the photometry. Of the 13 stars considered with SSPP and SPHINX metallicity estimates, 11 (85\%) fall within the 1$\sigma$ region. The remaining two stars each lie within the 1.5$\sigma$ region. Both the median and $C_{\rm BI}$ computed for the residuals were found to be $-0.02$\,dex. These 13 sources with spectroscopic parameters are provided in Table~\ref{table:M15}. We use the SP subscript to denote values determined by SPHINX, while SS refers to spectroscopically determined estimates from the SSPP.

\begin{figure*}
	\centering
	\includegraphics[trim={3.50cm 0.0cm 3.cm 0.00cm},clip, width=\textwidth]{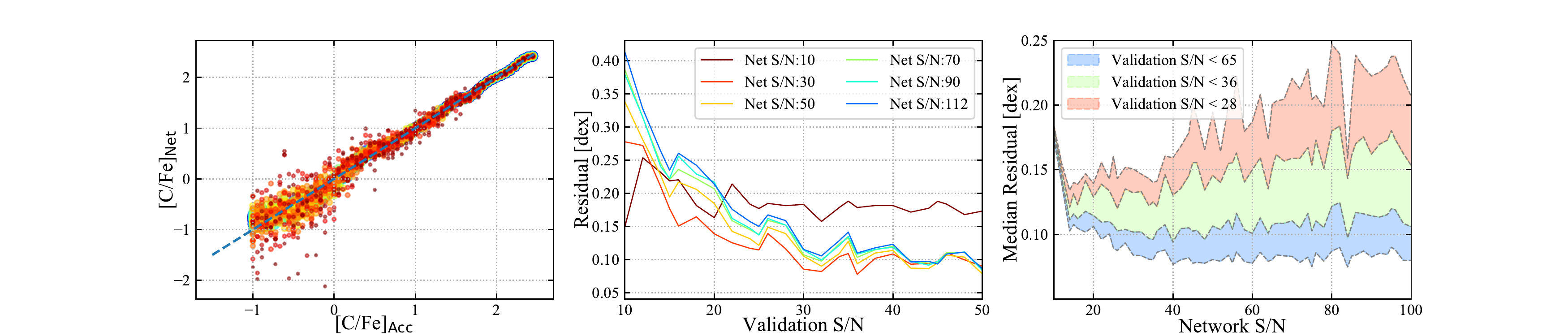}
	\caption{\textit{Left panel}: [C/Fe] estimates from the network trained on $S/N$ = 122 inputs; the color of the estimates are proportional to the signal-to-noise of the validation photometry, where red indicates lower $S/N$. \textit{Middle panel}: [C/Fe] residuals for networks of various signal-to-noise are shown, as a function of the $S/N$ for the validation sets employed. \textit{Right panel}: Median residuals for three tiers of validation sets, $S/N < 28$, $S/N < 36$, and $S/N < 65$, are shown, as a function of the network $S/N$.}
	\label{fig:carbon_result}
\end{figure*}

J-PLUS photometry of M15 provides an opportunity to validate the effective temperature and metallicity routines on a single stellar population. While determinations of metallicity made with SPHINX exhibit a scatter of $\sim$0.26\,dex, we are able to make estimations of central metallicity for M15 to remarkably high accuracy, in addition to estimates of photometric effective temperature that are commensurate with estimates from the SSPP. In fact, the uncertainty in the mean of the metallicity distribution generated from estimates using J-PLUS photometry is within the uncertainty in the value determined from high-resolution spectra. While the capability of SPHINX is currently limited with respect to individual precision measurements, SPHINX enables studies of stellar populations as well.

%-------------- CARBONICITY SECTION --------------
\section{Preliminary results on carbonicity}\label{carbon_section}

While SPHINX is presently limited to obtaining effective temperature and metallicity determinations, we anticipate the addition of carbonicity ([C/Fe]) and surface gravity ($\log{g}$) estimates in the near future. This will require more stars with J-PLUS photometry for proper training on their carbon features. However, with our library of synthetic spectra, we are able to explore the validity of a neural network approach to carbon detection. The application of the synthetic library also permits an in-depth study of the influence of underlying signal-to-noise on network training and estimation.

To isolate the carbon dependence on our synthetic spectra, we restricted our analysis to synthetic spectra corresponding to a giant VMP star, $T_{\rm eff}$ = 5000\,K, $\log{g}$ = 1.00, [Fe/H] = $-$2.50. Next, 100 spectra were generated by way of cubic spline interpolation across the interval $-$1.0 $<$ [C/Fe] $<$ +2.5. These normalized spectra were convolved with a blackbody to best approximate a flux-calibrated spectrum.

\par The flux of each spectrum was randomly rescaled from 50\% to 150\% to best emulate a random distribution of distance moduli, then injected with Gaussian noise, resulting in 66 batches of the original 100 spectra from 10 $<$ $S/N$ $<$ 122, each batch consisting of a unique global signal-to-noise. Synthetic magnitudes for the narrow-band J-PLUS filters were computed in the same manner as for the SDSS Reservoir, by convolving the noise-injected, blackbody-calibrated synthetic spectrum with the appropriate filter-response function. Linear scaling was then performed to center each input magnitude distribution.

For simplicity, SPHINX was limited to a single ANN unit for each trial described below. For each trial, one of the 66 synthetic spectra batches was used to train the ANN unit, consisting of six narrow-band inputs -- $J0378$, $J0395$, $J0430$, $J0515$, $J0660$, and $J0861$ -- with a single hidden layer of five hyperbolic tangent neurons and a stochastic gradient-descent optimization. The result was an array of identical networks, each trained on a set of inputs with a specific signal-to-noise, which we refer to as the network $S/N$. These networks of unique $S/N$ were then tested on each batch of noise-injected synthetic spectra, in order to evaluate the performance of each version of the network with an approximation of photometry of varying quality. In doing so, we assigned a median residual to each network and its constituent 66 verification batch runs. The result of this analysis is shown in Fig. \ref{fig:carbon_result}.

The left panel of Fig. \ref{fig:carbon_result} shows the results of the $S/N$ = 122 network for all batches of synthetic spectra. We find a characteristic behavior, namely, that photometry of lower signal-to-noise results in a greater dispersion of the predicted [C/Fe] about the true value. The dispersion is more prevalent in the lower carbonicity regime ([C/Fe] $<$ 0.0), as might be expected, since the carbon-sensitive features at low [C/Fe] are weaker, and so are generally more influenced by the $S/N$ across the spectrum.

The middle panel of Fig. \ref{fig:carbon_result} shows the behavior of each network, as a function of the $S/N$ in the photometry in the validation set. First, there is a general trend of a reduction in the spread of the residuals as the signal-to-noise of the validation photometry increases. Interestingly, in the input range of $S/N < 20$, networks trained on $S/N$ = 10 and $S/N$ = 20 outperform those trained on higher-quality photometry, for which residuals exceeded a standard deviation of 0.30\,dex.  We conclude that, in general, a network trained on poor photometry will exhibit better performance on poor photometry.

The right panel of Fig. \ref{fig:carbon_result} tracks the performance of the network for different regions of the validation signal-to-noise, as a function of the network's S/N. For sources of low quality ($S/N < 28$), the median residual is worse overall for all values of the network $S/N$. The low-quality source residuals tend to increase as the networks are trained on higher-quality photometry. In the region of 20 $\le$ $S/N$ $\le$ 30, we observe no significant increase in the median residual of the low-quality source set (median residual $\sim$ 0.13\,dex), while the high-quality set relaxes to a roughly consistent value of $\sim$ 0.08\,dex. This suggests an optimal region of $S/N$ for network training, approximately $30 < S/N < 40$, to ensure  
high-confidence performance for sources of both high and low quality, for network applications of this nature.

%%%%%%%%%%%%%%%%%%%%%%%%%%%%%%%%%%%%%%%%%%%%%%%%%%%%%%%%%

\section{Discussion and conclusion}\label{conclusion}

The Stellar Photometric Index Network Explorer, SPHINX, is designed to estimate the effective temperature and metallicity of stellar atmospheres using broad- and intermediate-band optical J-PLUS photometry, with a particular emphasis on the capacity to identify low-metallicity stars. This pipeline attempts to optimize training databases provided to converge its constituent artifical neural network units, as well as the relative weight assigned to these subordinate ANN units for use in science estimates. By doing so, SPHINX has the potential to be quite flexible in its ability to accommodate a variety of photometric data.

Estimates of effective temperature made using J-PLUS Early Data Release were found to be successful across a temperature range of 4500 $< T_{\rm eff}$ (K) $< 8500$, with an uncertainty of $\pm$ 91\,K. Comparisons were made to previous calibrations by \citet{Lee:2008a} and \citet{Fukugita:2009}, and in all cases estimates by SPHINX with J-PLUS photometry proved superior to the broad-band performance. For the application of both broad-band photometric methods, the J-PLUS analogs to SDSS photometry were used. We emphasize the success of temperature estimates made with SPHINX without the use of a-priori knowledge of surface gravity, for which we find no significant influence on temperature determinations within $2.0 < \log{g} < 5.0$.

\par Photometric estimates of metallicity for stars in the J-PLUS First Data Release within 4500 $< T_{\rm eff}$ (K) $<$ 6200 indicate sensitivity down to [Fe/H] $\sim$ $-$3.0, with a scatter of $\sigma$([Fe/H]) =0.25\,dex. However, verification of this sensitivity is limited by the lack of numerous [Fe/H] $< -$3.0 stars in the footprint of J-PLUS at present. We find that metallicity determinations made with J-PLUS DR1 photometry are not influenced by surface gravity within the range $1.75 < \log{g} < 5.0$. SPHINX is a very effective as a means for recovering and correctly identifying significant fractions of low-metallicity stars. SPHINX recovers $85 \pm 3$\,\% of very metal-poor (VMP) targets in the DR1 testing set used, with $63 \pm 4$\,\% of VMP candidates correctly classified. For the trial consisting of synthetic magnitudes from the SDSS Reservoir, $70 \pm 6$\,\% of stars designated as [Fe/H]$<-3$ were confirmed by the SSPP, while all remaining stars were VMP. Of the 43 EMP stars in the synthetic SDSS trial, $53 \pm 6$\,\% were recovered by SPHINX.

Photometric metallicity estimates were made using J-PLUS photometry of 664 sources associated with globular cluster M15 (NGC 7078). The central value obtained, [Fe/H]$=-2.32\pm 0.01$, with a residual spread of 0.29\,dex, was commensurate with the value determined from SEGUE-1 medium-resolution spectra of $-2.32\pm 0.02$ for the cluster. These estimates essentially match the accepted value from the literature, $-2.33\pm0.02$ \citep{Harris:1996,Carretta:2009}. 
The accuracy of estimates made by SPHINX was, in part, due to the level of optimization made possible by way of adjustable neural network architecture and the ability to tailor the training set used to converge the subordinate ANN units. This enables SPHINX to make parameter estimates for science cases involving stars of a single evolutionary stage, or to generalize to accommodate field stars in a variety of evolutionary stages across large regions of sky.

The development of SPHINX is at present limited to effective temperature and metallicity estimates, but we anticipate the addition of carbon abundance estimates, at least for some portion of the stellar parameter space. With carbon-abundance estimates from SPHINX, the primary sub-classes of carbon-enhanced metal-poor (CEMP) stars, the CEMP-$s$ and CEMP-no stars, describing the nature of their neutron-capture element abundance ratios \citep{Beers:2005}, could be discerned from photometry alone \citep{Yoon:2016}, avoiding the requirement for obtaining far-more time-consuming high-resolution spectroscopic follow-up. In addition, with accurate parallaxes from {\em Gaia DR2} \citep{Gaia:2016}, improved photometric maps of carbon abundances for the inner- and outer-halo regions of the Milky Way are readily obtainable (see spectroscopic-based maps in \citealt{Lee:2013}).

In the future, we plan to provide photometric parameters for a substantial portion of stars in the J-PLUS footprint. However, at present, a pure catalog of stellar sources from J-PLUS is not yet available -- present catalogs are  contaminated by quasars at faint magnitudes. Low-redshift quasars can be identified and removed by straightforward PSF analysis and the use of color-color diagrams \citep{Caballero:2008}. However, high-redshift quasars remain a challenge, as they are optically similar to ultracool dwarfs.

The artificial neural network methodology of SPHINX, in conjunction with J-PLUS photometry, is an ideal tool for selecting low-metallicity targets for spectroscopic follow-up. In addition, reasonable sensitivity has been demonstrated for individual estimates of metallicity down to [Fe/H] $\sim -3.0$ for stars in the range $4500 <T_{\rm eff}$ (K)~$ <6200$. We anticipate that future data releases from J-PLUS will enable an expanded sensitivity to low-metallicity stars. Ultimately, we expect to substantially increase the number of known VMP, EMP, UMP, and CEMP stars in the coming years on the basis of such studies.

\begin{acknowledgements}
	D.D.W., V.M.P., and T.C.B. acknowledge partial support from Grant No. PHY-1430152 (Physics Frontier Center/JINA Center for the Evolution of the Elements), awarded by the U.S. National Science Foundation.

%%% GLOBAL J-PLUS
Based on observations made with the (JAST/T80 and/or JST/T250) telescope/s at the Observatorio Astrof\'{i}sico de Javalambre, in Teruel, owned, managed, and operated by the Centro de Estudios de F\'{i}sica del Cosmos de Arag\'{o}n.
We thank the OAJ Data Processing and Archiving Unit (UPAD) for reducing and calibrating the OAJ data used in this work.

Funding for the J-PLUS Project has been provided by the Governments of Spain and Aragon through the Fondo de Inversiones de Teruel, the Spanish Ministry of Economy and Competitiveness (MINECO; under grants AYA2016-79425-C3-2-P, AYA2015-66211-C2-1-P, AYA2015-66211-C2-2, AYA2012-30789 and ICTS-2009-14), and European FEDER funding (FCDD10-4E-867, FCDD13-4E-2685). The Brazilian agencies FAPESP and the National Observatory of Brazil have also contributed to this project.

% Stavros Akras
S.A. acknowledges support of CNPq, through  postdoctoral  fellowship  (project  300336/2016-0).

MVCD thanks the FAPESP scholarship process number 2014/18632-6.

S.R. acknowledge financial support from CAPES, CNPq, and FAPESP.

T.M. acknowledges support provided by the Spanish Ministry of Economy  and Competitiveness (MINECO), under grant AYA-2017-88254-P.

R.L.O. was partially supported by the Brazilian agency CNPq (Universal Grants 459553/2014-3, PQ 302037/2015-2, and PDE 200289/2017-9).

CMO and LSJ acknowledge the support from FAPESP and CNPq.

%Young Sun
Y.S.L.acknowledges support from the National
Research Foundation (NFR) of Korea grant funded by the Ministry of Science
and ICT (No.2017R1A5A1070354 and NRF-2018R1A2B6003961).

%Dupke
RAD acknowledges support from CNPq through BP grant 312307/2015-2, CSIC through
grant COOPB20263, FINEP grants REF. 1217/13 - 01.13.0279.00 and REF 0859/10 - 01.10.0663.00 for partial hardware support for the J-PLUS project through the National Observatory of Brazil.

%Paula Coelho
PC acknowledges support from CNPq 305066/2015-3.

%%%%% SDSS acknowledgement
Funding for the Sloan Digital Sky Survey IV has been provided by the Alfred P. Sloan Foundation, the U.S. Department of Energy Office of Science, and the Participating Institutions. SDSS-IV acknowledges
support and resources from the Center for High-Performance Computing at
the University of Utah. The SDSS web site is www.sdss.org.

SDSS-IV is managed by the Astrophysical Research Consortium for the 
Participating Institutions of the SDSS Collaboration including the 
Brazilian Participation Group, the Carnegie Institution for Science, 
Carnegie Mellon University, the Chilean Participation Group, the French Participation Group, Harvard-Smithsonian Center for Astrophysics, 
Instituto de Astrof\'isica de Canarias, The Johns Hopkins University, 
Kavli Institute for the Physics and Mathematics of the Universe (IPMU) / 
University of Tokyo, Lawrence Berkeley National Laboratory, 
Leibniz Institut f\"ur Astrophysik Potsdam (AIP),  
Max-Planck-Institut f\"ur Astronomie (MPIA Heidelberg), 
Max-Planck-Institut f\"ur Astrophysik (MPA Garching), 
Max-Planck-Institut f\"ur Extraterrestrische Physik (MPE), 
National Astronomical Observatories of China, New Mexico State University, 
New York University, University of Notre Dame, 
Observat\'ario Nacional / MCTI, The Ohio State University, 
Pennsylvania State University, Shanghai Astronomical Observatory, 
United Kingdom Participation Group,
Universidad Nacional Aut\'onoma de M\'exico, University of Arizona, 
University of Colorado Boulder, University of Oxford, University of Portsmouth, 
University of Utah, University of Virginia, University of Washington, University of Wisconsin, 
Vanderbilt University, and Yale University.

\end{acknowledgements}

%-------------------------------------------------------------------
\bibliographystyle{aa}
\bibliography{mainBIB.bib}

\clearpage

\end{document}

%% file: M15_latex_table.dat
21:29:37.1 & 12:13:41 & 5570 & 93 & 5447 & 75 & -2.44 & 0.24 & -2.47 & 0.24 \\ 
21:29:39.4 & 12:18:21 & 5411 & 84 & 5299 & 44 & -2.34 & 0.26 & -2.38 & 0.10 \\ 
21:29:43.1 & 12:00:18 & 5690 & 55 & 5521 & 60 & -2.21 & 0.17 & -2.25 & 0.13 \\ 
21:29:48.5 & 12:20:20 & 5510 & 81 & 5315 & 37 & -2.37 & 0.28 & -2.27 & 0.16 \\ 
21:29:49.9 & 12:18:12 & 5288 & 93 & 5206 & 48 & -2.42 & 0.25 & -2.34 & 0.17 \\
21:29:52.3 & 12:19:40 & 5128 & 108 & 5215 & 48 & -2.24 & 0.63 & -2.27 & 0.20 \\ 
21:30:00.7 & 12:22:32 & 5568 & 136 & 5320 & 33 & -2.35 & 0.27 & -2.27 & 0.16 \\ 
21:30:13.5 & 12:00:38 & 5523 & 91 & 5390 & 36 & -2.42 & 0.22 & -2.33 & 0.18 \\ 
21:30:15.2 & 12:19:46 & 5352 & 114 & 5269 & 50 & -2.21 & 0.26 & -2.33 & 0.18 \\ 
21:30:24.5 & 12:02:38 & 5592 & 130 & 5420 & 33 & -2.95 & 0.79 & -2.35 & 0.18 \\ 
21:30:35.2 & 12:18:10 & 5626 & 117 & 5314 & 32 & -2.36 & 0.26 & -2.49 & 0.17 \\
21:30:49.3 & 12:07:31 & 5532 & 114 & 5344 & 31 & -2.52 & 0.26 & -2.32 & 0.14 \\ 
21:30:54.4 & 12:07:11 & 5569 & 121 & 5419 & 45 & -2.39 & 0.22 & -2.23 & 0.21 \\ 